\documentclass[%
 reprint,
 aps,
 prx,
longbibliography,
]{revtex4-2}

\usepackage{booktabs}
\usepackage[utf8]{inputenc}
\usepackage{bbm}
\usepackage{appendix}

\usepackage[usenames,dvipsnames]{xcolor}
\usepackage{color}
\usepackage{colortbl}

\definecolor{qctrl_primary}{HTML}{680Ce9}
\definecolor{qctrl_secondary}{HTML}{BF04DC}
\definecolor{qctrl_noise}{HTML}{7B7479}
\definecolor{qctrl_axis_labels}{HTML}{514B4F}
\definecolor{qctrl_borders}{HTML}{CFCBCE}
\definecolor{qctrl_blue}{HTML}{4177D8}
\definecolor{qctrl_aqua}{HTML}{32A4A8}
\definecolor{qctrl_green}{HTML}{32A857}
\definecolor{qctrl_lime_green}{HTML}{A2A933}
\definecolor{qctrl_orange}{HTML}{D6742F}
\definecolor{qctrl_red}{HTML}{D84144}
\definecolor{qctrl_fuchsia}{HTML}{D84190}

\usepackage{blindtext}

\usepackage{amssymb,amsmath,amsthm}
\usepackage{mathtools} 

\usepackage{mathrsfs}
\usepackage{dsfont} 
\usepackage[nice]{nicefrac}

\usepackage{cancel}
\usepackage{bm}

\RequirePackage{algorithm}
\RequirePackage{algorithmic}

\theoremstyle{definition}

\theoremstyle{remark}

\usepackage{physics}
\usepackage{siunitx} 
\usepackage{qcircuit}
\usepackage{braket}

\usepackage{graphicx}
\usepackage{float}
\usepackage{rotating}

\usepackage{dcolumn}
\usepackage{multirow}
\usepackage{makecell}
\usepackage{bigdelim}
\usepackage{blkarray}
\usepackage{array}

\usepackage{hyperref}
\hypersetup{
final=true,
plainpages=false,
pdfstartview=FitV,
pdftoolbar=true,
pdfmenubar=true,
bookmarksopen=true,
bookmarksnumbered=true,
breaklinks=true,
linktocpage,
colorlinks=true,
linkcolor=blue,
filecolor=blue,      
urlcolor=cyan,
citecolor=blue,
anchorcolor=green
}

\usepackage[nameinlink]{cleveref}
\crefname{equation}{Eq.}{Eqs.}
\crefname{align}{Eq.}{Eqs.}
\crefname{figure}{Fig.}{Figs.}
\crefname{table}{Table}{Tables}
\crefname{tabular}{Table}{Tables}
\crefname{section}{Sec.}{Secs.}
\crefname{appendix}{App.}{Apps.}

\crefname{appsec}{App.}{Apps.}
\crefname{appchapter}{App.}{Apps.}

\crefname{algorithm}{Algo.}{Algos.}
\creflabelformat{equation}{#2#1#3}

\usepackage{diagbox}
\begin{document}

\title{Experimental benchmarking of an automated deterministic error suppression workflow for quantum algorithms}
\author{Pranav S. Mundada}
\thanks{These two authors contributed equally}
\author{Aaron Barbosa}
\thanks{These two authors contributed equally}
\author{Smarak Maity}
\author{Yulun Wang}
\author{Thomas Merkh}
\author{T. M. Stace}
\author{Felicity Nielson}
\author{Andre R. R. Carvalho}
\author{Michael Hush}
\author{Michael J. Biercuk}
\altaffiliation[Also ]{ARC Centre for Engineered Quantum Systems, The University of Sydney, NSW Australia}
\author{Yuval Baum}

\affiliation{%
 Q-CTRL, 
 Sydney, NSW Australia \& Los Angeles, CA USA \& Berlin, Germany
}%

\date{\today}


\begin{abstract}
Excitement about the promise of quantum computers is tempered by the reality that the hardware remains exceptionally fragile and error-prone, forming a bottleneck in the development of novel applications. 
In this manuscript, we describe and experimentally test a fully autonomous workflow designed to deterministically suppress errors in quantum algorithms from the gate level through to circuit execution and measurement. 
We introduce the key elements of this workflow, delivered as a software package called {\sl Fire Opal}, and survey the underlying physical concepts: error-aware compilation, automated system-wide gate optimization, automated dynamical decoupling embedding for circuit-level error cancellation, and calibration-efficient measurement-error mitigation. 
We then present a comprehensive suite of performance benchmarks executed on IBM hardware, demonstrating up to $>1000\times$ improvement over the best alternative expert-configured techniques available in the open literature. 
Benchmarking includes experiments using up to 16 qubit systems executing: Bernstein Vazirani, Quantum Fourier Transform, Grover's Search, QAOA, VQE, Syndrome extraction on a five-qubit Quantum Error Correction code, and Quantum Volume. 
Experiments reveal a strong contribution of Non-Markovian errors to baseline algorithmic performance; in all cases the deterministic error-suppression workflow delivers the highest performance and approaches incoherent error bounds without the need for any additional sampling or randomization overhead, while maintaining compatibility with all additional probabilistic error suppression techniques.  

\end{abstract}

\maketitle



\section{Introduction}
Large-scale fault-tolerant quantum computers are likely to enable new solutions for problems known to be hard for classical computers. 
The usefulness of a quantum computer is ultimately determined by its ability to successfully implement meaningful and relevant quantum algorithms in reasonable times and with reasonable resources. 
While in recent years demonstrations by Google~\cite{AruteNature2019} and the Chinese Academy of Sciences~\cite{Zhong1460} showed the first steps toward quantum advantage, reliable implementation of generic quantum algorithms is still out of reach.  In order to characterize and predict the performance of emerging quantum machines, a variety of system-level benchmarking strategies has been devised, including the creation of comprehensive system-level metrics such as Quantum Volume~\cite{Cross19}, and the development of algorithmic benchmarking test suites~\cite{Lubinski21}.
Demonstrating the ability to run mid-scale algorithms on NISQ devices with high success probability is an essential step in the route to quantum advantage in the near future, and large-scale fault-tolerant quantum computing in the long term. 

In augmenting the performance of NISQ-era algorithms, much attention has focused on performance optimization at the level of individual quantum logic gates, pushing them towards nominal fault-tolerance thresholds~\cite{Motzoi2009, Baum2021, Boixo_RL,an2019, NiuNPJ2019, Yale_Optimal,carvalho2020errorrobust,soare2014, werninghaus2020leakage,Wittler2021}. 
However, experimental demonstrations~\cite{Proctor17, Proctor22} have identified a gap between the algorithmic performance projected from gate-level proxy measures such as randomized benchmarking and the actual performance achieved on real hardware. 
This reflects the fact that the quality of a quantum algorithm is affected not only by the quality of the individual constituent components (qubits, gates, measurements), but also by the interplay of the global device and algorithmic properties such as device topology, multi-qubit noise correlations, and circuit structures.  Benchmarking results on commercial and research machines \cite{tomesh2022supermarq} presented to date consistently reveal that errors arising from multiple sources substantially degrade performance and limit the utility of contemporary hardware.

In this work, we introduce an automated workflow for deterministic error suppression~\cite{Fire_patent} in quantum algorithms and present a wide range of performance benchmarks highlighting how algorithms with varying characteristics can be augmented through this tooling. 
We perform experimental demonstrations using a software package called {\sl Fire Opal} on commercial hardware from IBM and demonstrate $>1000\times$ improvement over the best expert-configured implementations using tools available from the respective platforms.   We present results for comprehensive algorithmic benchmarking covering the breadth of techniques studied in the literature, including the following algorithms exhibiting widely varying characteristics (connectivity, circuit depth, etc): Bernstein-Vazirani (BV), quantum Fourier transform (QFT), Grover Search, Quantum Approximate Optimization Algorithm (QAOA), Variational Quantum Eigensolver (VQE), quantum error correction (QEC), and Quantum Volume (QV).  For each algorithm, we present a detailed performance evaluation and describe in detail the relevant quantitative metrics used.
In our analyses we demonstrate that there are typically substantial performance gaps between the performance predicted from proxy measures such as gate-level randomized benchmarking and realized algorithmic performance; we ascribe these to non-Markovian error sources which are efficiently suppressed using deterministic error suppression \cite{White20, White21,White22}. 
Across devices, topologies, and algorithms we show that through a holistic view of error suppression from the gate level to the circuit level we can approach incoherent error limits. 
Importantly, these techniques are \emph{deterministic} and suppress errors with no additional user overhead in circuit execution, in contrast with sampling techniques which involve many repeated executions of each circuit \cite{Wallman16,Hashim21,Minev_PEC, Kandala19}. Further, by presenting direct demonstrations combining this workflow with QEC in order to improve syndrome extraction, we highlight the complementarity of low-level error suppression and logical encoding on the path to large-scale fault tolerance. 

We begin with an overview of typical hardware performance and link to key error channels observed on NISQ-era devices, many of which deviate substantially from the assumption of Markovian errors. 
We then introduce the key elements of an automated, deterministic error suppression pipeline designed to maximally suppress hardware errors to incoherent limits. 
This is followed by full reporting of performance benchmarking with quantitative analysis of observed performance enhancement across a range of test algorithms. 
We conclude with a brief discussion and future outlook.

\section{Background and context}
\subsection{Performance of algorithms on typical hardware}
Quantum algorithms fail when executed on faulty hardware. This simple phenomenology is routinely observed in both commercial and laboratory settings through a variety of proxy measures such as circuit success probability, or the overlap of the measured output distribution relative to an ideal output.  Ultimately this phenomenology is linked to errors in the underlying hardware elements.

Most hardware backends provide tabulated data for the error rate associated with each gate in an algorithm, as measured using device-level characterization protocols such as randomized benchmarking, cycle benchmarking and gateset tomography \cite{Knill08, Magesan11, Blume13,Erhard19,corcoles13, Xia15,Barends14}. These protocols return average proxy measures which characterize the fidelity of single and multi-qubit gates, the building blocks of quantum algorithms.  With this information, one may calculate the expected performance of the algorithm by assigning the backend-provided error rates to each operation (and qubit) in the compiled machine instructions. 

Algorithmic-performance estimates derived from these proxy measures routinely overestimate actual hardware utility~\cite{Proctor17, Proctor22, Mavadia18}. This leads to a simple but fundamental observation: proxy measures for constituent operations are in isolation generally poor predictors for system-level performance. This is further reflected in research-community-based benchmarking results spanning multiple systems~\cite{Lubinski21}.  As shown later, the presence of noise and error correlations in space and time dominate this discrepancy and open opportunities for the deployment of new techniques for performance enhancement leveraging concepts from quantum coherent control.

\subsection{Review of errors sources in near-term quantum computers}
Overall, understanding and predicting algorithm-level performance requires consideration of a much wider variety of error processes than typically discussed in the theoretical literature.  In this subsection, we review the most common realistic error sources encountered in hardware, including non-Markovian (correlated or coherent) errors which violate the assumption that errors are statistically independent.

\emph{Single qubit coherence limits ($T_1$/$T_2$)}:
The quantum state of a single qubit can be represented by a complex superposition of its internal states, $A|0\rangle + B\exp{i\phi}|1\rangle$ with $A^2 + B^2 \leq 1$. While quantum gates intentionally steer quantum states towards a desired target, quantum noise, and unwanted couplings may steer quantum states as well. For example, noise processes may arise due to \emph{energy loss} to the environment, in which a qubit state $|1\rangle $ decays to $|0\rangle$. This process is characterized by a typical time scale known as $T_1$ and represents a stochastic process producing statistically independent errors. 
In another process called \emph{dephasing}, the phase $\phi$ is randomized through interaction with the environment (for instance fluctuating magnetic fields). The underlying processes which cause dephasing are often characterized as being stochastic, but in reality, tend to exhibit strong temporal correlations which result in coherent errors. This process is characterized by a typical time scale known as $T_2$ which may be much shorter than $T_1$. 
This is the first canonically non-Markovian error process typically encountered in real hardware.

\emph{Calibration and gate errors}: 
Universal quantum computing can be achieved using a small and finite ``universal gate set" composed of single-qubit rotations and a multi-qubit coupling mechanism. Both single-qubit and two-qubit gates are typically realized by irradiating the qubits with shaped wave packets (pulses). During each operation, incoherent $T_{1}$ decay processes contribute a gate error approximately proportional to the ratio of $\tau_{g}/T_{1}$, with $\tau_{g}$ the gate duration.

Additional non-Markovian errors can arise from the imperfect execution of the gate itself. The ideal gate implementation has specific ideal pulse parameters (amplitude, duration, frequency, waveform) and any deviation of these parameters causes gate errors. Such deviation can arise due to an imperfect calibration process (e.g. identifying the qubit's native frequency), slow variations in system parameters, or signal distortions in generation or transmission.  Simultaneously, individual gate operations can experience non-Markovian processes such as leakage out of the qubit subspace induced by the driving control signal's spectral content.  All of these processes produce errors with long-time correlations across many gates in a circuit.

\begin{figure*}[t!] 
\centering
\includegraphics[width=\linewidth]{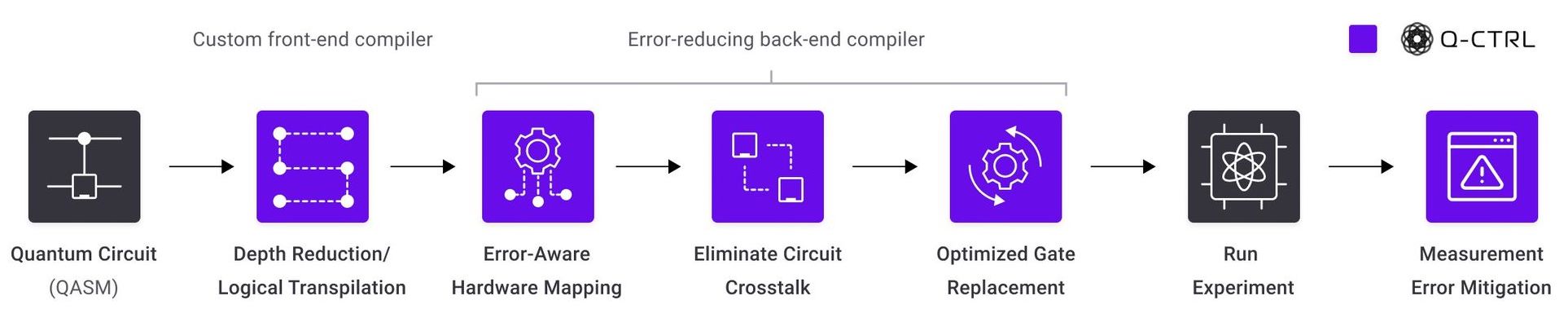}
\caption{The {\sl Fire Opal} automated error-suppressing workflow for quantum algorithms. Each step (described in the main text) is executed autonomously with no user configuration, given an arbitrary input circuit and  access to a supported quantum hardware backend. Measurement error mitigation is performed after the hardware backend returns raw results, and the output is returned to the user.}
\label{fig:Pipeline}
\end{figure*}

\emph{Circuit level crosstalk}:
In addition to errors impacting individual qubits, errors may also arise due to unwanted coherent coupling or semi-classical driving between nearby qubits, resulting in another substantive form of non-Markovian error. These phenomena are known as crosstalk which may arise from a wide range of physical mechanisms. Quantum crosstalk emerges due to spurious coupling between qubits via the mechanism of device fabrication or operation. The most common spurious coupling in superconducting qubits is an unwanted $ZZ$ coupling; in this process, a qubit acquires a deterministic phase shift that depends both on its state and a nearby qubit state. Classical crosstalk emerges due to the unwanted driving of a qubit by control signals applied to neighboring devices. Even when the drive on a neighboring device is off-resonant, weak coupling over long circuits can introduce unwanted rotation errors.  This phenomenon may actually appear nonlocal across a device because of the relatively long wavelength of microwave control signals employed in many architectures, and the challenges of microwave hygiene at the chip scale.  Both forms of crosstalk can produce substantial errors on idling qubits during the execution of a circuit;  classical crosstalk can also cause challenges in the implementation of parallel temporally synchronous gates within a circuit.

\emph{Measurement errors}:
Measurement in quantum computers involves a process of interaction with the quantum device causing a probabilistic collapse of the system's superposition state to one of the measurement basis states.  This physical process is itself susceptible to error from the device-level interrogation procedure.  For instance, the microwave radiation used in reading out superconducting circuits can induce unwanted transitions between states or the process of interpreting a (typically analog) readout signal such as the phase of reflected microwaves from a resonator can lead to imperfections in assigning a $\ket{0}$ on $\ket{1}$ state.  These individual readout errors grow exponentially with qubit count and are often spatially correlated, adding substantial complexity in disambiguating between various output bitstrings at the conclusion of a computation.

\section{Automated workflow for deterministic error suppression}\label{Sec:workflow}
We introduce an efficient protocol for error suppression that is designed to enhance the performance of quantum algorithms on quantum hardware.  Our strategy is based on purely deterministic approaches that reduce errors without the use of sampling or randomization methods; i.e. they do not require additional execution overhead in repetition in order to attain error reduction.

In this workflow, the user provides an algorithm specified in the QASM intermediate representation \cite{Cross17} and our protocol executes a series of deterministic error suppression strategies to augment the machine instructions prior to execution. Additional post-processing is then employed in order to improve readout efficiency.

In Appendix~\ref{App:C} we describe the detailed methodologies in use and their relative contributions to performance, but here provide a brief summary of the error-reduction pipeline, which is schematically visualized in Fig.~\ref{fig:Pipeline}.  
\begin{enumerate}
 
    \item \emph{Depth reduction and logical transpilation} -- A sequence of compiler passes is used to mathematically reduce the depth (gate count) of the quantum circuit and the logical operations in the circuit are mapped to the native gates available on the hardware. This step is executed in runtime on user initiation of the pipeline.
    \item \emph{Error-aware hardware mapping} -- Error-aware compilation is used to best select the appropriate subset and logical assignment of qubits on a device. For any circuit with a width less than the total number of qubits on the device, there is a combinatorially large range of logically equivalent circuits.  Additional compilation passes select and rank the best implementations among these, accounting for device topology, tabulated gate errors, parallel-gate crosstalk, etc. This step is executed in runtime on user initiation of the pipeline.
    \item \emph{Eliminate circuit crosstalk} -- Dynamical decoupling (DD) sequences are incorporated to mitigate various idling errors including dephasing and $ZZ$ crosstalk at the algorithmic level. The underlying pulse sequences are optimized for the specifications of a particular backend via an automatic ``ranking" routine. The ranking routine is context-aware, applying optimal sequences to each qubit's idle period with knowledge of single-qubit gate errors as well as the gate sequences applied to neighboring qubits in order to account for both quantum and classical (single-qubit-gate) crosstalk. This step is executed in runtime on user initiation of the pipeline.
    
    \item \emph{Optimized gate replacement} -- Backend error rates and device coherence times are consumed in order to identify gates with substantial headroom above incoherent ($T_{1}$) limits. Parallelized AI-powered optimizers then automatically perform an efficient analog-level gate optimization.  The process involves automated parsing of the device topology to ensure parallel gate optimizations do not share qubits; multiple parallelized steps are executed in series to ensure all relevant single and or multi-qubit gates are optimized. These routines are performed in the background and are not part of the user workflow.  The output of this optimization is a lookup table of new analog-layer gate definitions that are called at runtime in defining the overall circuit.  
    
    \item \emph{Run experiment}. The user circuit is configured for execution in an appropriate language supporting both gate timing and analog-layer waveform controls such as qiskit pulse, pyquil, or AWS braket \cite{IBM_website, Amazon_website,  Rigetti_website}. The circuit is then executed on the hardware backend over a specified number of shots and results returned.  In this manuscript, we focus on demonstrations employing IBM hardware with a corresponding choice of low-level language.
    
    \item \emph{Measurement-error mitigation} -- A final post-processing step is performed on all shots returned from the backend.  An initial AI-driven calibration routine identifies measurement-error processes, and this calibration data is stored. 
    In runtime, the calibration data and measurement results are combined in order to estimate the probability distribution of the hardware's output bit strings and an associated confidence interval. 
    This process scales as $\mathcal{O}(1)$ with qubit count, providing an efficient mechanism to capture correlations with low calibration overhead and fast processing.  This step is executed in runtime with no additional overhead on the part of the user.
  
\end{enumerate}

\begin{table*}[tp!]
\centering
\begin{tabular}{|l| >\raggedright p{5.75cm} | p{5cm} | p{4cm} |} 

 \hline
 \bf Algorithm & \bf Characteristics & \bf Evaluation metric & \bf Performance enhancement \\ 
 \hline\hline

  \vtop{\hbox{\strut Bernstein-}\hbox{\strut Vazirani}} & Shallow and Sequential - circuit depth scales linearly with circuit width (for full connectivity) and native entangling gates do not overlap.

  & Success probability of returning the all ‘1’ state. This selection constitutes a worst-case scenario using the most challenging (state-dependent) oracle.
  
  & $>1000\times$ improvement in the success probability, up to 16 qubits.\\
 \hline
 \vtop{\hbox{\strut Quantum Fourier}\hbox{\strut Transform}\hbox{\strut (QFT)}}
& Intermediate depth and partially sequential - circuit depth scales quadratically with circuit width (for full connectivity). Parts of the algorithm can be parallelized.
 & Success probability averaged over 10 initial states (chosen to be inverse QFT of single bitstrings).
 & $>100\times$ over default (expert), up to 12 qubits.\\
 \hline
 
 Grover Search & High depth, scaling depends on the implementation of the multi-CX gate. The depth can be reduced by allowing the addition of ancilla qubits.
 & Selectivity and distance (circuit infidelity) to the ideal target probability distribution over 32 different 5Q target states.
 & 8$\times$ improvement in circuit fidelity; transform selectivity ($S<0 \to S>1$) for all target states, up to 5 qubits.  \\
 \hline
 
\vtop{\hbox{\strut Quantum}\hbox{\strut Approximate}\hbox{\strut Optimization}\hbox{\strut Algorithm}}
& Depth and density depend on the specific problem. Problem under test is a seven-qubit max-cut type Hamiltonian with depth and density similar to the QFT class.
 
 &  Structural similarity metric between the ideal and measured cost landscapes. 
 & 163$\times$ improvement in the Structural similarity metric.\\
 \hline

 \vtop{\hbox{\strut Variational}\hbox{\strut Quantum }\hbox{\strut Eigensolver}} & Depth and density depend on the specific ansatz. Problem under test uses four and six qubit $R_y$ ansatzes with linear entanglement. Each round of the ansatz has depth and density in the BV class.
 
 &  Ground state energy accuracy for BeH$_{2}$ molecule and deviation from the ideal set of Pauli expectation values using the Pearson distance. 
 & 5$\times$ error reduction in the mean ground state energy prediction, and 16$\times$ improvement in the Pearson distance of the measured Pauli expectation values. \\
 \hline
 \vtop{\hbox{\strut Quantum Error}\hbox{\strut Correction codes}}
 &  Depth in the QFT class. Due to device topology, the encoding block and each of the syndrome inference blocks are highly sequential. & Agreement between syndrome readout and errors inferred from direct data qubits readout. &  4$\times$ improved error detection in the repetition code and $3.3\times$ for the full five qubit code (total 9 qubits).\\ 
  \hline
  
 Quantum Volume & Intermediate depth (similar to QFT) with maximal density. Operation are maximally parallelized. 
 
 & Mean heavy output (HO) probability, averaged over 300 random circuits sampled with 1000 shots each.
 & QV increased from 32 to 64.\\ 
 \hline

\end{tabular}
\caption{Benchmark algorithms. The algorithms chosen differ in their depth-to-width ratio, operations density, duration of idling periods and the sparseness of the ideal output distribution. 
These different characteristics pose different challenges and sensitivities to errors.}
\label{tab:Benchmark_algo}
\end{table*}

This error-suppressing pipeline is applicable to any quantum computing architecture, and is not specific to a particular qubit type; hardware-specific elements are captured by the various front-end and back-end compiler stages. 
Importantly, it can be executed independently in NISQ-era algorithms and also in the augmentation of fault-tolerant quantum error correction routines as demonstrated below. 
Further context on existing strategies for error management in NISQ era devices is provided in Appendix~\ref{Sec:AlternateStrategies} \cite{tripathi2022,Santos05, Sekatski16, Pokharel18, Wallman16,Temme17, Li17, Hashim21, Cai19, Minev_PEC, Kandala19}.

In practice, this workflow is ``zeroconfig" requiring no user intervention, configuration, or coding.  This stands in contrast to other techniques where substantial coding and configuration are required by a user to execute or combine various error-suppression strategies (see Appendix~\ref{Sec:AlternateStrategies}).  All calibration and optimization steps for measurement-error mitigation and automated gate optimization are executed in advance and results are stored for on-demand algorithmic execution.  Other protocols use runtime decision-making to select strategies based on the updated backend hardware parameters.  In practice, a user inputs a QASM definition of a circuit; execution and result-return is conducted with a single command.

\section{Algorithmic benchmarking with deterministic error suppression}\label{results}

\subsection{Benchmarking settings}
We now present a series of experimental demonstrations of algorithmic benchmarking using the error-suppressing pipeline described above. In order to understand performance advantages under widely varying settings -- locality, circuit depth, and width, circuit density -- we select a comprehensive suite of benchmarks, following previous publications on algorithmic benchmarking, augmented with protocols including quantum error correction and quantum volume.  
An overview of the selected benchmarks, their characteristics, and the underlying evaluation metrics is presented in Table \ref{tab:Benchmark_algo}.

We benchmark our methods against two software configurations and execute on several different IBM backends (See Table~\ref{tab:Benchmark_conf}). 
All algorithms are configured in Qiskit and then compiled to hardware instructions using the two alternative methods. 
The first method uses the default settings defined by the hardware provider and includes the most aggressive compilation scheme the platform provides. Most common users do not alter or add to the default settings. 
The second method we benchmark against uses the best expert-configured implementations of tools for error suppression available from the respective platforms; such available methods include interleaved dynamic decoupling, extra compilation strategies, and measurement-error~\cite{Bravyi21,Nation21}, which can be added by well-informed users. 
We refer to the second method as ``expert settings" as configuration requires a user to be knowledgeable about both the underlying strategy and how to configure the technique in circuit execution. 
When performing comparisons we ensure all configurations (Default, Expert, and Q-CTRL) are executed back-to-back in the same session in order to minimize the impact of daily parameter variations known to occur on these devices~\cite{Carvalho2021}.  Full details of the relevant benchmarking configurations are presented in table~\ref{tab:Benchmark_conf}.

\begin{table*}
\centering
\begin{tabular}{|p{2.5cm}| p{4.7cm} | p{5cm}|p{5cm}|} 
 \hline
 \bf Component & \bf Default & \bf Expert & \bf Q-CTRL  \\ 
 \hline\hline
 
  Compilation method & Qiskit level 3 Sabre compiler, best out of four seeds.  & Qiskit level 3 Sabre compiler combined with Qiskit mapomatic function for layout selection. &  Q-CTRL compilation passes and layout-selection optimization.\\
 \hline

 Quantum Logic Gates & Default & Default & Q-CTRL autonomously optimized two-qubit gates.\\
 \hline
 
 Error mitigation strategies  & None & Qiskit native dynamical decoupling and Qiskit measurement-error mitigation (``complete" for circuits with $\leq7$ qubits and ``M3" $>7$ qubits). & Q-CTRL optimal context-aware embedding of crosstalk-suppressing dynamical decoupling sequences and Q-CTRL scalable measurement-error mitigation.\\
 \hline
\end{tabular}
\caption{Benchmarking configurations as implemented on the IBM backend. Devices used: Lagos, Jakarta and Guadalupe.}
\label{tab:Benchmark_conf}
\end{table*}

\begin{figure*}[t!] 
\centering
\includegraphics[width=\linewidth]{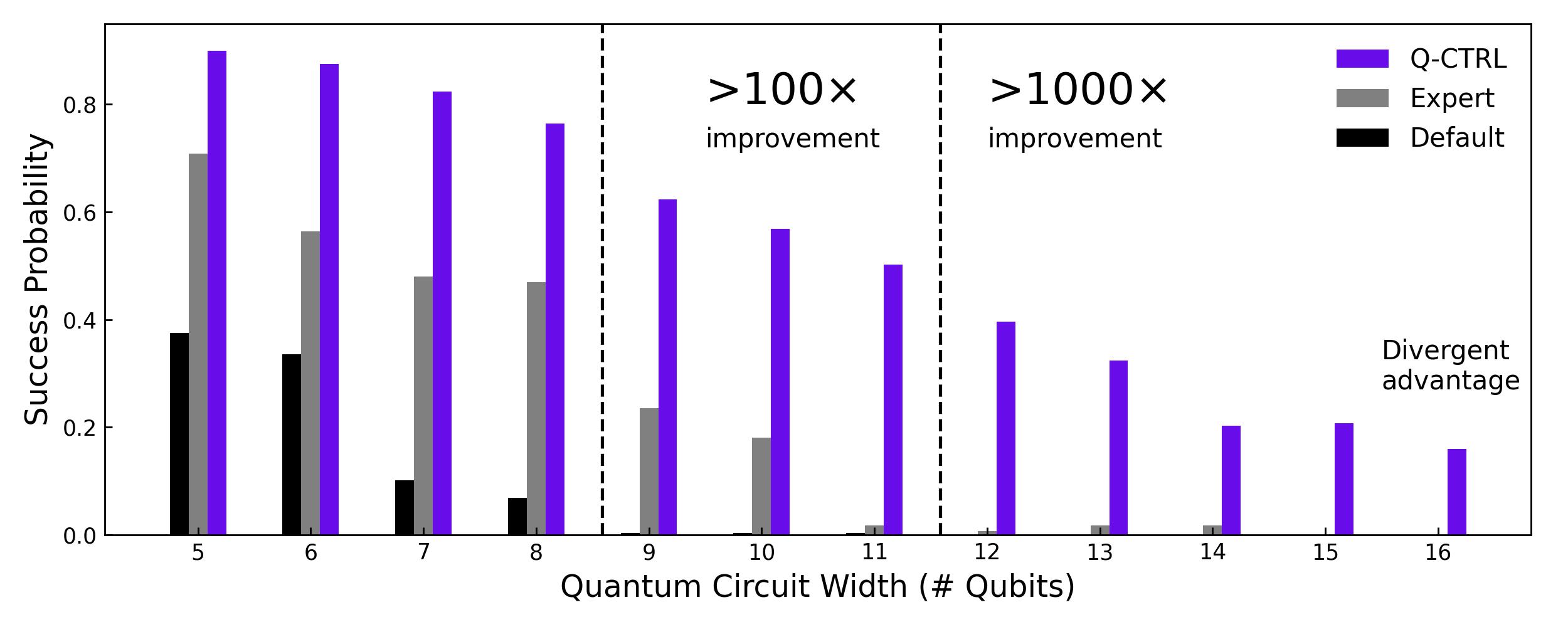}
\caption{Success probability of a BV algorithm as a function of the number of qubits. Default, expert, and Q-CTRL settings are described in~\Cref{tab:Benchmark_conf}. Associating the mode of the distribution with the correct answer, the Q-CTRL pipeline returns the correct answer for all system sizes while the other methods fail to return the correct answer beyond 8 qubits (default) and beyond 10 qubits (expert).}
\label{fig:BV}
\end{figure*}

\subsection{Deterministic algorithms}
Deterministic algorithms are designed to perform a specific task. Given an initial state, their ideal outcome if fully deterministic. Such algorithms include, among others, the Bernstein-Vaziranialgorithm, the quantum Fourier transform, Grover's search, Shor's algorithm, Hamiltonian evolution, and more.
As some of these algorithms require thousands of entangling gates even for a small number of qubits, we focus our attention on the BV and QFT algorithms along with an optimized version of a five-qubit Grover's search.

\subsubsection{Bernstein-Vazirani}
An $N$-qubits BV algorithm has a deterministic answer in the form of a single $(N-1)$-qubits bitstring. The answer (target) bitstring is determined by the choice of oracle circuit. We choose the oracle circuit in such a way that the answer is given by the all '1' bitstring. This oracle is the most challenging one among the possible BV oracles as its circuit includes the highest number of entangling gates, hence, the results we obtain can be viewed as the worst performance among all possible oracles.
As the ideal answer is a single bitstring, we use \emph{success probability} (SP) as a quality measure, i.e., the probability to measure the correct bitstring. 

Typical comparative performance for BV is illustrated in Fig.~\ref{fig:BV}. 
For both default and expert setting, as the number of qubits grows, the likelihood that the algorithm returns the correct answer diminishes. 
By the time the algorithm integrates just nine qubits, the likelihood of success for the default configuration is less than approximately one percent. 
For 16 qubits, in this demonstration, the base hardware does not return the correct answer over 32,000 attempts. 
Adding the deterministic error-suppressing pipeline delivers substantial benefits for all circuit widths; as the width increases, the relative performance advantage increases with qubit count. 
In this demonstration at 16-qubit circuit-width, the Q-CTRL pipeline delivers a success probability of approximately $\sim20\%$, with daily performance fluctuations bringing this occasionally near $40\%$. 
In all trials executed, the Q-CTRL error-suppressing pipeline always delivers the best performance for all circuit widths.

\begin{figure}[t!] 
\centering
\includegraphics[width=\linewidth]{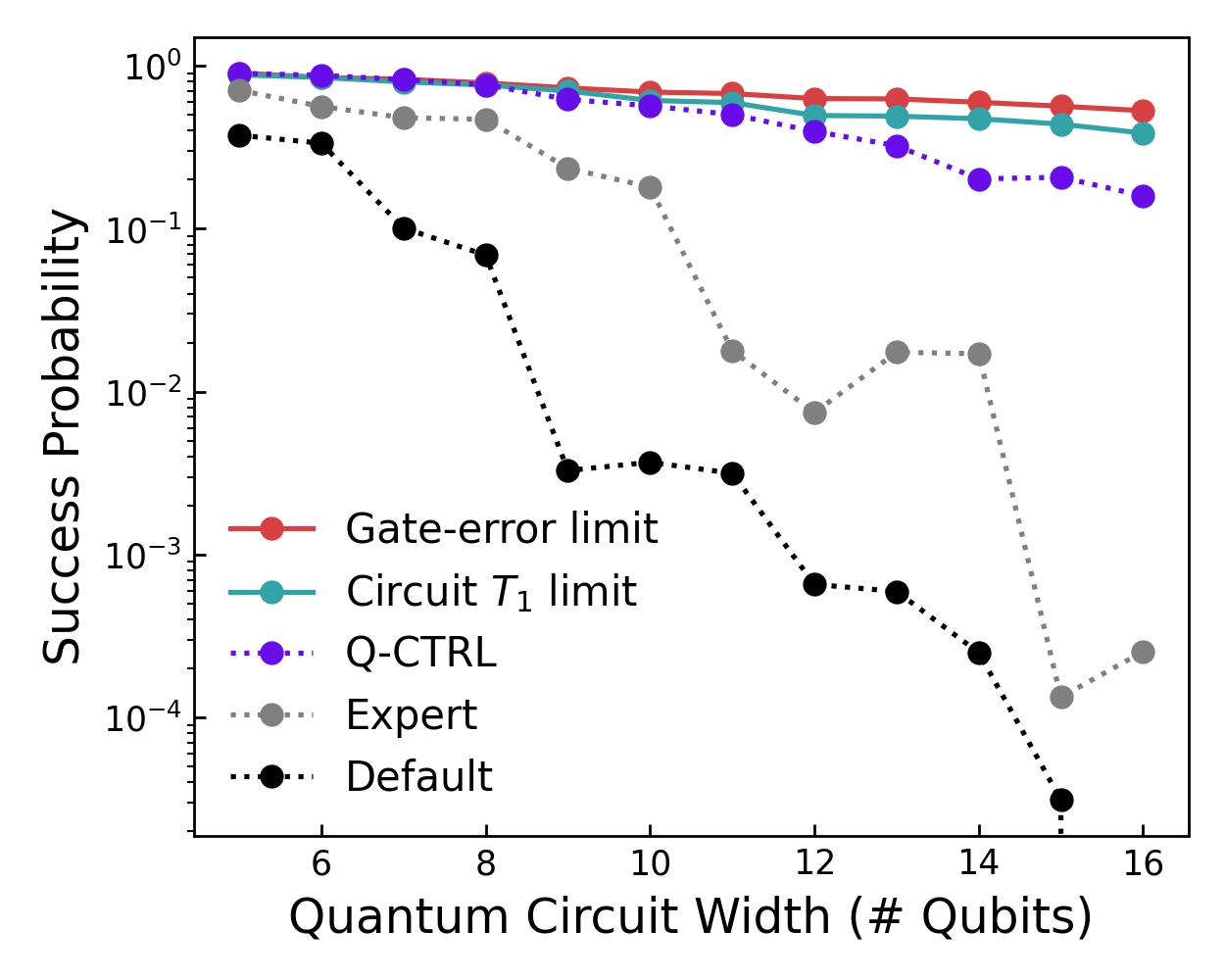}
\caption{Success probability of the BV algorithm for the different settings as appeared in Fig.~\ref{fig:BV}, together with
two projected performance bounds which assume the existence of either uncorrelated gate errors or $T_1$ processes only. 
The gate-error limit is calculated by a simple multiplication of the fidelities derived from the backend for all single and two-qubit gates in the circuits.
The circuit $T_1$ limit is calculated by a multiplication of the $T_1$-limited fidelity of each participating qubit, calculated as $\prod_{i} \exp(-T_{i}/2T^{(i)}_1)$; here $T_{i}$ is the total active time of qubit $i$ (from first operation until measurement).
}
\label{fig:BV_gap}
\end{figure}

Further detail is revealed when examining these results on a logarithmic scale and comparing the achieved success probability to that predicted using simple proxy error measures, see Fig.~\ref{fig:BV_gap}. 
The first proxy measure uses backend-provided gate-error rates in order to estimate the circuit fidelity given the count of various gate operations (Gate-error limit), assuming gate imperfections are the only source of error. 
A second proxy measure is estimated assuming performance is bounded by irreversible $T_{1}$ decay processes occurring throughout the execution of the circuit. 
The two estimates are similar, but the $T_{1}$ limit in this case is slightly lower due to the presence of idle periods which are not properly accounted assuming only gate errors.

We observe, consistent with Refs~\cite{Proctor17, Proctor22}, that the achieved performance using the default and expert settings is up to five orders of magnitude lower than the bounds derived from simple proxy measures. 
This observation highlights the importance of correlated error processes that are not typically captured effectively by considering independent-gate-error models. 
By contrast, the Q-CTRL pipeline delivers performance within a factor of order unity of the two proxy bounds. 
Additional observations show that over an eight-month period, the results from the Q-CTRL pipeline remain constant within a factor of order unity, while for any particular circuit width, the performance of the default and expert configurations can vary up to over an order of magnitude (see Appendix~\ref{App:Stability}). 
There remains a residual opportunity for improvement, but these comparisons give strong evidence that deterministic error suppression can effectively stabilize against correlated error processes that otherwise limit algorithmic performance.

\subsubsection{Quantum Fourier Transform}
The quantum Fourier transform represents a mapping between quantum states in the Hilbert space. In order to efficiently estimate its performance without the need for full multi-qubit state tomography, we initialize a circuit to a state which is an inverse Fourier transform of a single bitstring.  Such states can be generated with high fidelity by a single layer of Hadamard gates followed by a single layer of virtual $Z$ rotations; these still form a full superposition of all possible bitstrings despite the simple initialization process. After execution of the QFT algorithm on these initial states, we measure the probability of finding a specific bitstring as the output of the algorithm.
We repeat this procedure for different initial states and average the success probability achieved for each target output bitstring. 

\begin{figure}[t!] 
\centering
\includegraphics[width=\linewidth]{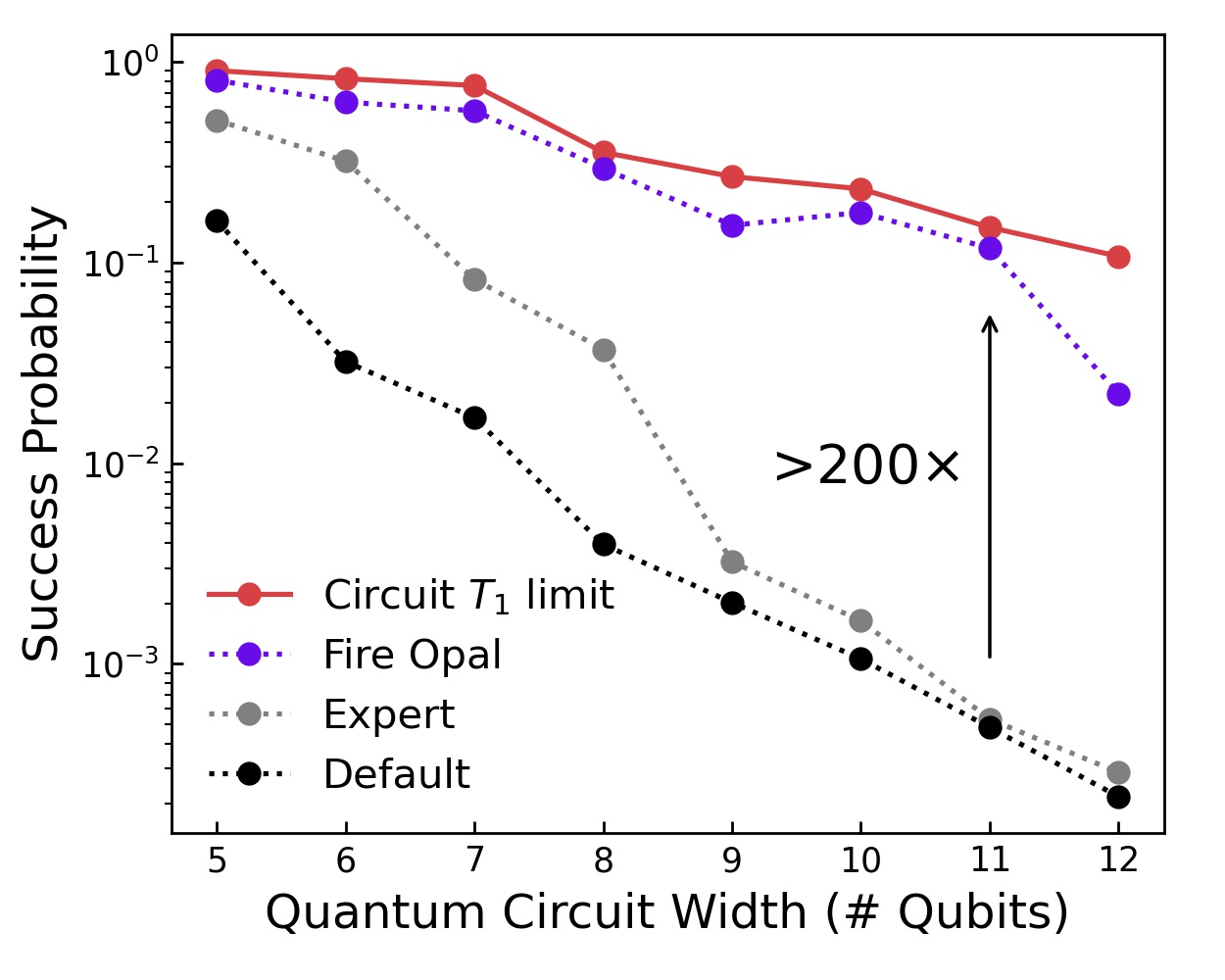}
\caption{The mean success probability, over 10 different initial states, of QFT circuits as a function of the system size. Absolute performance is lower than for BV due to the deeper circuit structure.}
\label{fig:QFT}
\end{figure}

As in the BV demonstrations, we observe rapid degradation of measured success probability with circuit width for the default and expert settings when executed on hardware, see Fig.~\ref{fig:QFT}. In this case, the greater circuit depth used for QFT results in a more rapid decay than BV. The addition of the Q-CTRL error-suppressing pipeline shows a consistent improvement up to $200\times$ over both the expert settings and default settings.

\begin{figure*}[t!] 
\centering
\includegraphics[width=\linewidth]{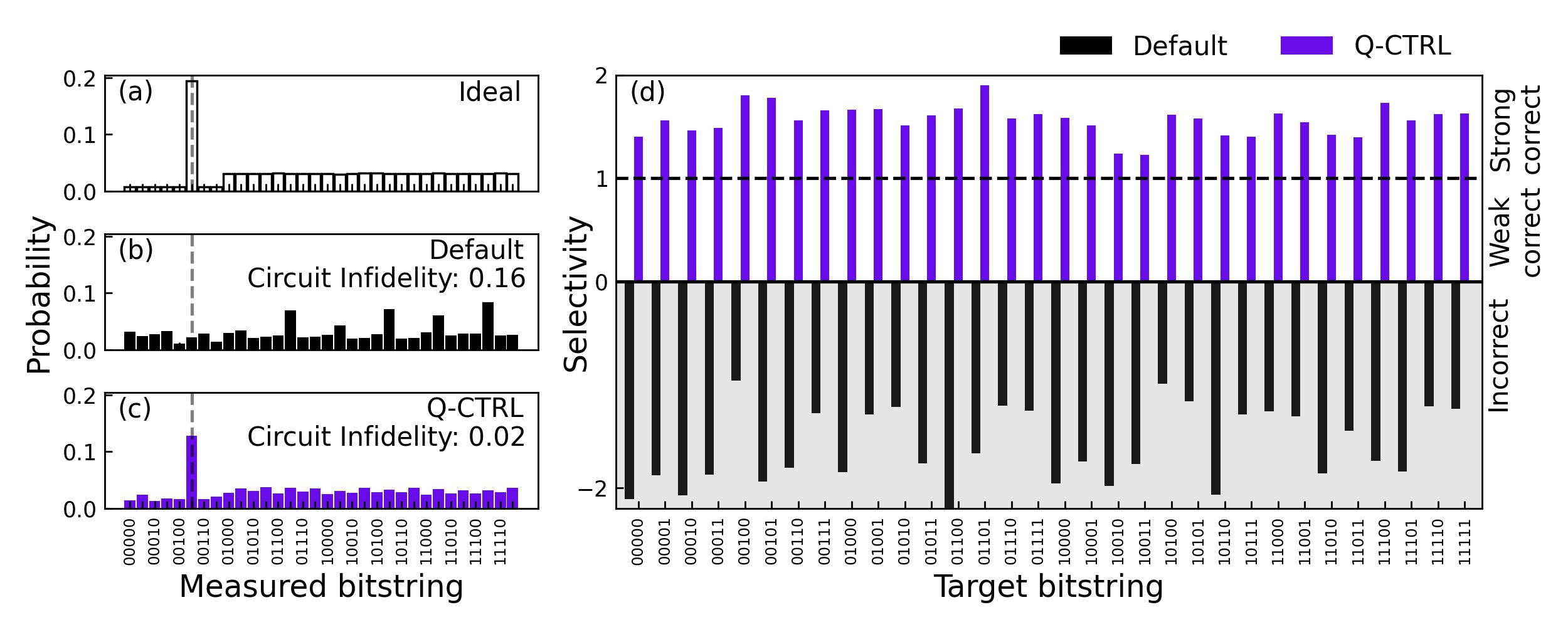}
\caption{(a-c) The output probability distributions over all possible 5Q bitstrings after two iterations of a Grover search algorithm with a specific oracle (00101 as a solution). (a) is the ideal expected distribution as calculated using a noiseless simulator, and (b-c) are probabilities obtained using default settings (black) and using the Q-CTRL pipeline (purple). 
The resemblance between the measured and ideal distributions is captured by the circuit infidelity. The distribution obtained by using Q-CTRL tools is $8\times$ closer to the ideal results. Similar improvement is observed for all target states. (d) Selectivity calculated for all target bitstrings, demonstrating $S>1$ for all possible 5 qubit states using the Q-CTRL pipeline compared to $S<0$ using the default settings.}
\label{fig:Grover}
\end{figure*}

\subsubsection{Grover's Search}
The last deterministic algorithm we benchmark is Grover's search, whose solution contains a predefined set of bitstrings (we focus on the case where the solution set includes a single bitstring). The solution of the search is encoded as the mode of the output distribution; we compare the measured distribution over all possible bitstrings to the ideal distribution in order to determine success probability (see Table~\ref{tab:Benchmark_algo}). We quantify the agreement between the ideal and the measured distributions using \emph{circuit infidelity}, $I = 1 - (1-H^2)^2$, with $H$ being the Hellinger distance between the two probability distributions.

In order to quantify the usefulness of the search, we define the selectivity, $S = \log_2{\left(\frac{p_t}{p_n}\right)}$, where $p_t$ is the probability of measuring the target bitstring of the search and $p_n$ is the probability of measuring the most probable bitstring which is not the target of the search. 
The number of shots needed in order to guarantee the correct solution with high confidence level scales as $p_t^{-1}S^{-2}$, i.e., $p_t$ alone is not a sufficient measure of the quality of a Grover's search or any algorithm that encodes the solution in the mode of the output distribution. 
Effectively, this metric enables a quantitative measure of the algorithm's ability to discriminate the target state from a noisy background distribution. 

Large positive values of $S$ indicate that the correct answer can be obtained consistently and efficiently (Strong correct regime), i.e., by sampling the algorithm a small number of times (in the limit $S\to\infty$ a single shot is guaranteed to return the correct answer). Low, yet positive, values of $S\leq 1$ allow finding the correct answer consistently but require additional sampling overhead in order to yield the correct answer (Weak correct regime).  For $S\leq0$, independent of the number of samples, the algorithm will consistently yield an incorrect answer (incorrect regime). 

We present the results of a five-qubit Grover's benchmark in Fig.~\ref{fig:Grover}.   We implement an optimized construction of the search algorithm (which uses two additional ancilla qubits as presented in \cite{BeIt21}), in order to overcome the extreme circuit depth required in a device with weak connectivity. We restrict ourselves to implementing at most two iterations of the search protocol, approximately consistent with the expected $\sqrt N$ iteration for an $N$-qubit search required to guarantee convergence in the five-qubit case we test.  

In this benchmark, we consistently observe a binary distinction where the default implementation fails to yield the correct answer while the Q-CTRL pipeline correctly identifies the target.  In our observations, the default and expert configurations were qualitatively similar in their inability to produce a correct outcome from the search.  Examining more closely, a calculation of the selectivity for all five-qubit target bitstrings, shown in Fig.~\ref{fig:Grover}d, indicates that the default settings consistently result in $S<0$, while with the Q-CTRL error-suppressing pipeline $S>1$ in all cases.

\subsection{Hybrid algorithms}
Hybrid algorithms such as VQE and QAOA are increasingly popular due to their utility in a range of optimization problems and the potential to deliver quantum advantage in the NISQ era. They have been applied to a wide range of tasks from chemistry to transport and financial optimization. Despite their popularity, they are more challenging to benchmark than their deterministic cousins due to their hybrid quantum-classical structure.  The achieved performance in this case depends not only on the quantum processor performance, but also the choice of ansatz, the selected level of approximation, the number of circuit evaluations available, and the ability of the extrinsic classical optimization algorithm to navigate the quantum search space.  

We isolate QPU performance and explore the ability of our error-suppressing pipeline to improve the execution of the underlying quantum circuits in a single, arbitrary QAOA or VQE execution.
\subsubsection{QAOA}
We choose a typical problem Hamiltonian ($\mathcal{H}_p$)
for QAOA, a seven nodes and 10 edges weighted MaxCut, and fixing the approximation level (p). We re-parameterize the QAOA parameters in the Fourier basis reducing the number of parameters to two~\cite{Zhou2020}. We establish baseline performance by calculating the cost function (expectation value of $\mathcal{H}_p$) for large choices of parameters using an ideal (noise-free) simulator. We then execute the same circuits on hardware, directly evaluating the output for the same range of QAOA parameters and comparing the measured and the ideal cost landscape; in the limit of perfect QPU-hardware execution, these two landscapes should be identical. 

In order to quantitatively evaluate the similarity of the cost landscapes we use a structural similarity metric (SSIM) which is widely used in image analysis~\cite{Zhou04}. 
This metric compares two images in a manner known to be more robust to high-spatial-frequency noise than typical variance calculations. 
A score of one corresponds to ideal matching, and in this case, implies that the ideal and measured cost landscapes are identical. 

\begin{figure}[t!] 
\centering
\includegraphics[width=\linewidth]{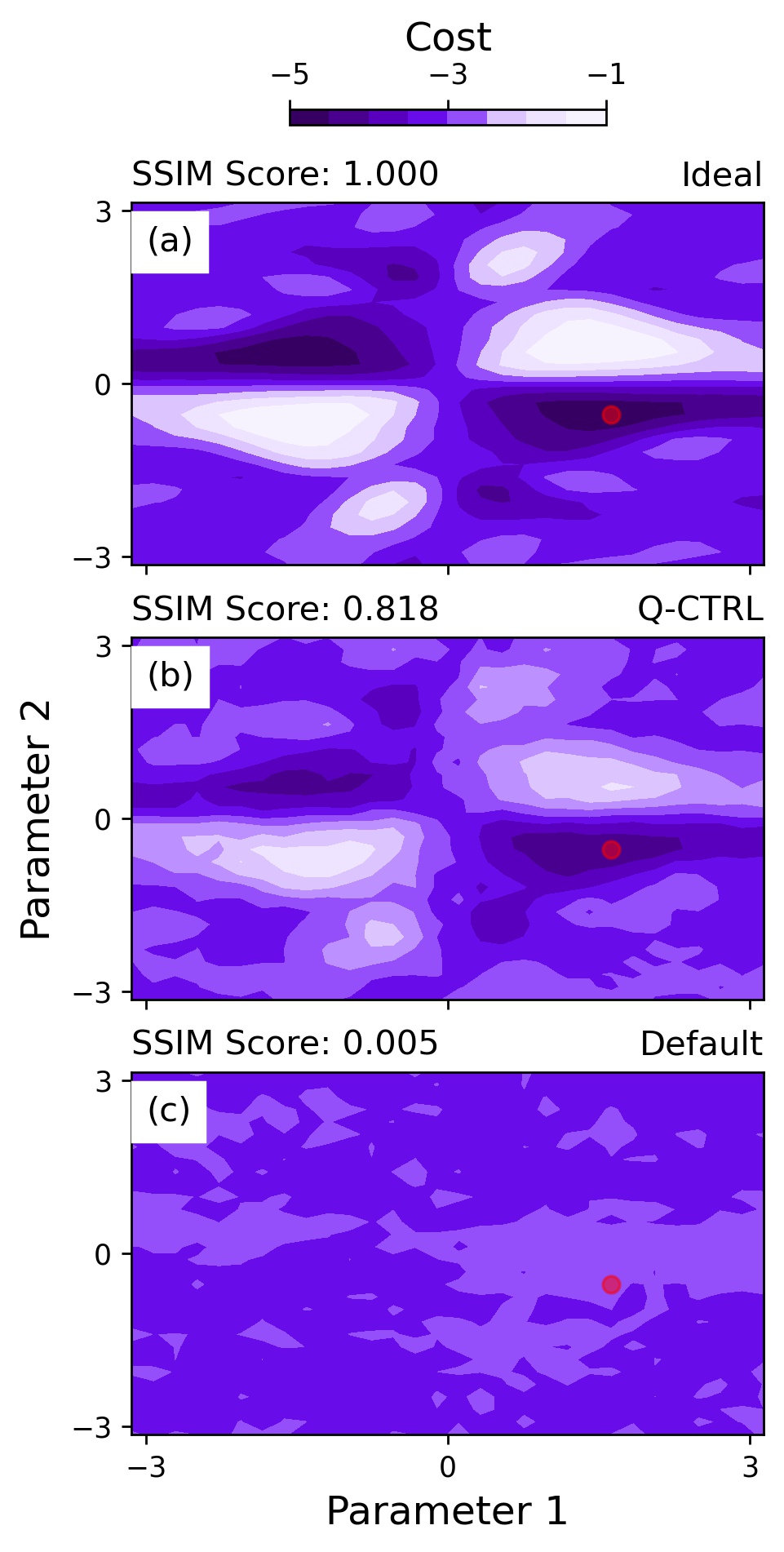}
\caption{The cost (expectation of H) landscape of a $p=3$ QAOA circuit representing a seven nodes and 10 edges weighted MaxCut problem. Here, we compare the ideal landscape (top), the experimentally obtained landscapes using the Q-CTRL pipeline (middle) and using the default settings (bottom). The ideal optimum is denoted by the red dot. Structural similarity relative to the ideal landscape is calculated and shown for each setting (see text).}
\label{fig:QAOA}
\end{figure}

As an example we calculate the SSIM comparing the default hardware implementation against the ideal landscape, and demonstrate SSIM$\sim0.005$, indicating low structural similarity to the ideal landscape shown in Fig.~\ref{fig:QAOA}. 
In the presence of realistic, strongly correlated hardware errors the QAOA landscape exhibits areas of relatively barren plateaus with respect to the ideal landscape.  Adding the Q-CTRL pipeline increases the SSIM by $163\times$ to SSIM$\sim0.818$. 
This demonstrated ability of deterministic error suppression at the hardware level to improve QPU performance in QAOA stands in contrast to sampling routines postulated to be unable to improve the influence of barren plateaus~\cite{Wang21}.

\subsubsection{VQE}
The variational quantum eigensolver uses the variational principle to compute the ground state energy of a Hamiltonian. VQE does so through the use of a parameterized circuit with a fixed form known as a \emph{variational form} or an Ansatz.

We benchmark the performance of VQE by utilizing it to compute the ground state energy of Beryllium hydride ($\text{BeH}_2$) molecule \cite{Kandala17}, which can be well described (with a minimal loss of precision) by considering four interacting atomic orbitals and a spin degree of freedom, which in turn, can be mapped to a 6-qubit problem. Such a mapping converts the Hamiltonian of the molecule into a sum of multi-qubit Pauli operators over the 6 qubits.
Further approximations can be made (freezing additional degrees of freedom) leading to a 4-qubit representation of the molecule's Hamiltonian. Both the 4- and 6-qubit representations of the Hamiltonian are taken from the Qiskit-nature library \cite{Qiskit}.

We generate trial wavefunctions using the two-local ansatz with two repetitions, in which $R_y$ gates are used to generate rotations and CNOT gates are used to generate entanglement. This variational form is known as the $R_y$ ansatz with linear entanglement, and it has 18 free parameters \cite{Tilly21}.

We evaluate the ground state energy of the $\text{BeH}_2$ molecule for different bond distances ranging from 0.5\AA\ to 5\AA.  For each setting, we optimize the parameters by running the algorithm on an ideal simulator and then use the optimized parameters to compute the expectation of the Hamiltonian on real QPU hardware. We collect $8\times 10^3$ samples for each quantum circuit we run on the real QPU.

\begin{figure}[ht!] 
\centering
    \includegraphics[width=\linewidth]{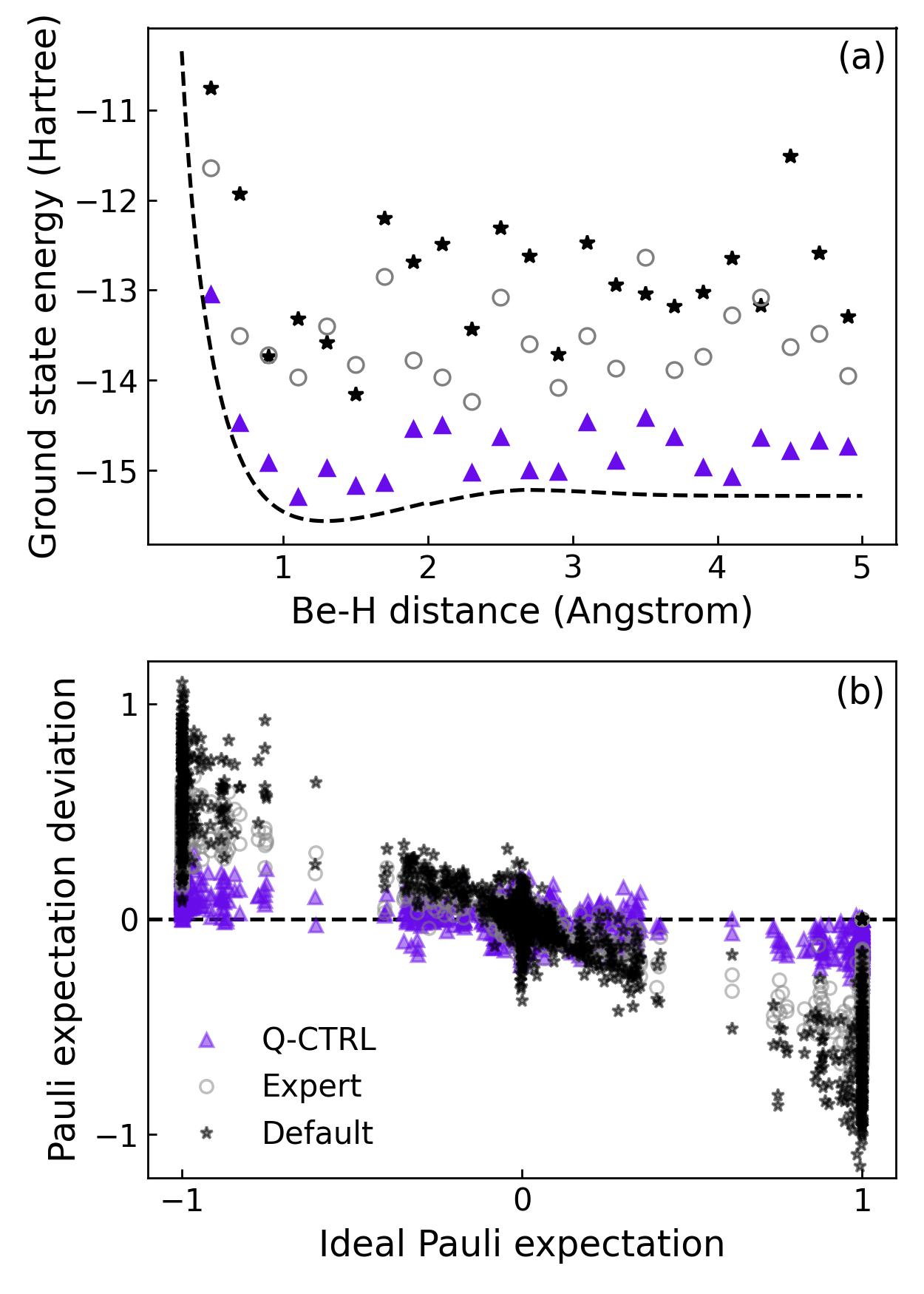}\caption{VQE performance with different workflows.  (a) Computed ground state energy of the $\text{BeH}_2$ molecule as a function of the Be-H bond distance for a 6-qubit ansatz. The dotted line displays results from an ideal simulator, and the individual points represent the results computed on real hardware with different settings. (b) Deviation from expected Pauli measurement as a function of the ideal value. Perfect agreement between theory and experiment would correspond to all points on the zero dashed line. To quantify the agreement, we calculate the Pearson distance, $P_d$ for the three methods - Default: 0.177, Expert: 0.042, Q-CTRL: 0.011}
\label{fig:VQE}
\end{figure}

Fig.~\ref{fig:VQE} shows the computed ground state energy of the $\text{BeH}_2$ molecule as a function of the Be-H bond distance for the 6-qubit ansatz.
The dotted line displays the result from an ideal simulator, and the individual points represent the results computed on real hardware with and without our pipeline;
inclusion of the Q-CTRL error-suppressing pipeline shows substantially better agreement between the estimated ground state energy and the ideal one. The mean energy deviation from the ideal value obtained using the Q-CTRL pipeline is about $5\times$ smaller compared to the hardware default for the 6-qubit ansatz. A similar experiment using the Q-CTRL pipeline with the 4-qubit ansatz (not shown) yields an energy deviation about $3.5\times$ smaller.

Each data point in Fig.~\ref{fig:VQE} (a) is obtained by a weighted sum of different Pauli expectation values. 
In Fig.~\ref{fig:VQE} (b), we show all the measured expectation values across different bond distances and compare them to their ideal values. We  quantify the agreement between the ideal and measured data using the Pearson distance, $P_d = 1 - P_c$, with $P_c$ being the standard Pearson correlation coefficient. A value of $P_d=0$ corresponds to perfect correlation where the deviation from the ideal value is zero for all points, while $P_d = 1$ corresponds to random data where the measured data is a random number between $-1$ to $1$ independently of the ideal value.
For six qubits, the agreement between the calculated expectation values and the ideal values is $16\times$  higher when using the Q-CTRL pipeline ($4\times$ higher with respect to the expert settings). 

We can further analyze the deviation from the ideal expectation values. As apparent, with default and expert settings, when the ideal expectation value is near zero, the deviation from the measured values is relatively small. This effect is expected, as random errors typically yield flat probability distributions that are characterized by small expectation values. In stark contrast, probability distributions that lead to extremal expectation values (near $\pm 1$), are unlikely to be generated by random noise. Indeed, both with default and expert settings, the deviation in expectation values near the extremal values grows dramatically as a function of the expectation value. With Q-CTRL settings, the deviation curve is approximately flat indicating that the expectation values are consistently originating from the correct probability distributions.   

Statistical methods, such as zero-noise extrapolation, Pauli twirling and random compilation (see Appendix \ref{Sec:AlternateStrategies}) can follow our deterministic pipeline to further improve the accuracy of the expectation values by reducing the effect of Markovian errors at the price of extensive overhead.

\subsection{Quantum Error Correction}

\begin{figure*}[t] 
\centering
    \includegraphics[width=\linewidth]{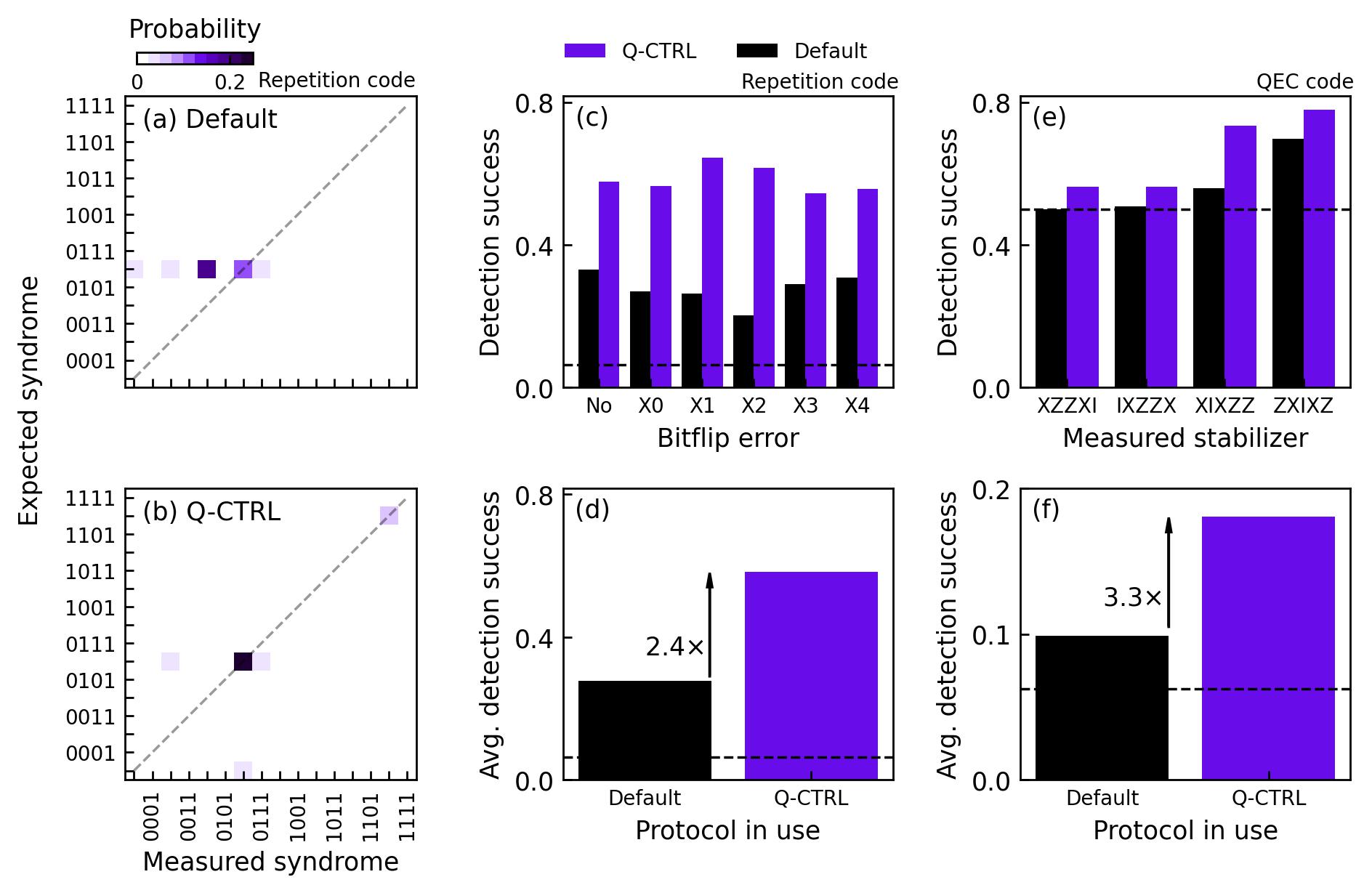}
\caption{Performance of QEC using the Q-CTRL pipeline.  (a-d) Benchmarking data for the five-qubit repetition code. (a-b) joint probability distribution of the measured and expected syndromes for the different settings after an error was injected to qubit 2.  The syndrome bit values ``0" or ``1" correspond to the measurement outcomes ``$+1$" and ``$-1$" for each of the stabilizer generators in the ordered set, $G$,  in  \cref{eqn:generators}. (c) Detection success as a function of the injected error for the default and Q-CTRL pipelines using color coding consistent with previous graphs.  Dashed line denotes the randomness threshold, which represents the success achieved by a random chance. 
(d) Mean detection success averaged over the six bit-error-location possibilities, comparing the two protocols. Arrow indicates quantitative performance enhancement, calculated with respect to the dashed line. (e-f) Data on full five-qubit QEC. (e) Detection success for each of the four stabilizers measured for the different pipelines. As before, the dashed line is the random chance value. (f) Overall mean (over injected errors) detection success incorporating data averaged over all four stabilizers simultaneously with net performance enhancement with respect to the random value represented. }
\label{fig:QEC}
\end{figure*}

Quantum error correction is a critical task that will be required for fault-tolerant quantum computation. 
At its heart, QEC depends on an \emph{encoding}, which redundantly maps  $k$ logical qubits into $n>k$ physical qubits, and $n-k$ \emph{stabiliser} generators, which provide partial information about hidden errors that may occur. 
Collectively, measurements of the stabilisers  produce \emph{syndrome} data, and a \emph{decoding algorithm}, processes the syndrome to provide a likely correction operator that has a high probability of correcting the error~\cite{nielsen00}.  

Different QEC codes have different performance characteristics, such as threshold \cite{dennis2002topological,PhysRevA.81.022317,Watson_2014,PhysRevX.9.041031}, geometry \cite{PhysRevLett.98.190504}, sparsity \cite{tillich2013quantum}, rate, and decoder complexity \cite{PhysRevLett.117.070501,PhysRevLett.127.040507}, all of which affect how effectively logical information can be stored. 
However, the basic physical processes are very similar across all stabiliser codes: ancillary qubits are entangled with the data qubits using circuits of one- and two-qubit gates to affect the few- or many-body stabiliser measurements that provide the error syndrome. 

At a low level, QEC is a form of a hybrid quantum algorithm that generates `mid-circuit' measurement results (i.e.\ syndrome data) used in a classical decoding process to infer and correct for errors during the course of a logical computation.  From this perspective, the algorithmic protocol constituting all key steps is amenable to circuit-level performance optimization using the tools described in Sec.~\ref{Sec:workflow}.

A common misconception is that QEC will render error suppression techniques obsolete. Since, all QEC schemes are essentially feedback-loop circuits that leverage qubits and operations overhead to ensure fault-tolerant computation, not only that QEC can not replace deterministic error suppression, but all QEC schemes will heavily rely on error-suppression techniques in order to lower the physical errors to a minimum. As the overhead required by QEC is exponential as a function of the native errors, no matter how low the native errors are, there is always a strong incentive to further reduce them.

We discuss two error-correcting codes: the 5-qubit repetition code which is capable of correcting up to two bit-flip errors, and the 5-qubit quantum error correction code, which is capable of correcting any single-qubit error. 
For each, we encode a logical qubit into the physical data qubits using a standard encoding circuit, inject an artificial error into the encoded qubit, and then perform syndrome measurements on ancilla qubits to indirectly read out the syndrome. 
Because of the limitations in mid-circuit measurement and conditional logic in the hardware interface we used, we were not able to perform real-time feedback for correction. 
We thus focus on the performance of a single iteration of the syndrome extraction protocol. 
We execute this process using both a default configuration and Q-CTRL's automatic error suppression pipeline, similar to the other benchmarking algorithms presented above.  

We first implement a circuit for encoding a logical $X_L$ state of the 5-qubit repetition code~\cite{nielsen00}, followed by one round of ancilla-assisted syndrome readout, and terminated by direct single-qubit measurements on all data and ancilla qubits; this circuit uses a total of nine qubits.  Because this code is classical, all the stabilisers, which are generated by the stabilizer generators in the ordered set:
\begin{equation}
    G=\{Z_{0}Z_{1}, Z_{1}Z_{2}, Z_{2}Z_{3}, Z_{3}Z_{4}\},\label{eqn:generators}
\end{equation}
which commute pointwise. 
As a result, the syndrome can be found either using ancilla qubits to give the \emph{measured syndrome} or determined directly using single-qubit measurements on the data qubits to give an \emph{expected syndrome}. 
Comparing the measured and expected syndrome, which ideally should agree, gives an estimate of the efficacy of the measurement.    

In \Cref{fig:QEC}(a)-(b) we apply an artificial bit-flip error on $q_2$, and measure the joint probability distribution of the measured and expected syndromes; perfect agreement of these distributions would yield only diagonal elements matching the location of any induced error.  Ideally, this error would generate the syndrome ``0110", and the probability of both measured and expected syndromes would be concentrated at this diagonal element. 
Imperfections manifest as off-diagonal elements with nonzero probability. 
Comparing the default and Q-CTRL pipelines, we see that the default configuration yields a joint probability distribution with high off-diagonal weight relative to the diagonal. 
This indicates that the syndrome extraction process fails more often than it correctly identifies the error location. 
By contrast, using the Q-CTRL protocol the joint distribution is much more constrained to the correct diagonal element at ``0110", indicating consistency between measured and expected syndromes, given the applied error. 
The error-detection success is measured as the trace over the joint probability distribution for each injected-error location in \Cref{fig:QEC}(c) and consistently shows higher performance for the Q-CTRL pipeline. 
We make quantitative comparisons between the two approaches by subtracting the contribution of random chance ($1/16$) from the detection success metric and calculating the performance ratio; we observe up to $4\times$ improvement for error-detection success with a single bit-flip error and an average of $2.4\times$ improvement over the five different error locations (\Cref{fig:QEC}(d)).

We next consider the performance of the 5-qubit QEC code, implemented using the encoding circuit from \cite{PhysRevApplied.15.034068}. 
In this case, stabilizers do not commute pointwise, and as a result, it is not possible to determine the full syndrome directly from single-qubit measurements at the end of the circuit. 
Instead, we can choose one of the stabilizer generators to measure pointwise, to extract the expected syndrome. 
For concreteness we consider the 5-qubit $[5,1,3]$ code which involves five data and four ancilla qubits used for syndrome measurement; the stabiliser generators are $\{XZZXI,IXZZX, XIXZZ, ZXIXZ\}$, which mutually commute but mix $X$ and $Z$ operators. 
We choose one of the generators to determine from single-qubit end-circuit measurements on the data qubits, asking whether the result agrees with the corresponding syndrome bit measured using ancilla qubits. 
Repeating this for different choices of the bit-wise stabilizer measurements, we  build up statistics for the success probability.

\Cref{fig:QEC}(e) shows the probability that a given direct measurement of a stabilizer generator matches the corresponding ancilla-measured syndrome bit for each of the four stabilizer generators, given all 15 possible single-qubit errors applied artificially. 
A probability of 50\% on this scale corresponds to completely random, uncorrelated outcomes; 100\% indicates that the measured syndrome bit and the directly determined stabilizer bit always agree. 
Results using the Q-CTRL protocol are consistently higher than the default pipeline, and all give an enhanced likelihood of error identification using the Q-CTRL pipeline. 
By contrast, two stabilizer measurements performed with default configurations are consistent with random chance. 
The average detection success per stabilizer above the randomness threshold (over the four stabilizers) is $\sim2.5\times$ higher using the Q-CTRL pipeline and the overall detection success (of all four stabilizers simultaneously) is $\sim3.3\times$ higher using the Q-CTRL pipeline (\Cref{fig:QEC}(f)).

We note that the hardware layout on which we implement these codes was optimized for the ``heavy-hexagon" QEC code \cite{PhysRevX.10.011022}, and was not optimal for the small demonstration codes we used. 
This contributes to the relatively low absolute performance of syndrome measurement obtained in our demonstrations. 
Nonetheless, we are able to compare the impact of hardware and circuit-level imperfections on the execution of the key algorithmic steps of QEC with our tests.

\subsection{Quantum Volume}
Quantum volume is a single metric that quantifies the largest random circuit of equal width and depth that a quantum computer can successfully implement. Higher-quality quantum computing systems are expected to have higher quantum volumes. Estimating quantum volume for hardware involves a complex protocol in which a wide range of random circuits are executed \cite{Cross19}. The ideal output distribution of each circuit is calculated using a noiseless simulation, and the heavy output set (all the bitstrings with higher probability than the median bitstring probability) is extracted. The circuits are then executed and sampled repeatedly in order to estimate the probability to measure a bitstring that belongs to the ideal set of heavy outputs. We refer to this probability as the HO probability. 

For each number of qubits, the HO probability is calculated and averaged over many random instances of the quantum volume circuits.  Any mean HO value above $2/3$ may be considered a success provided that either the standard deviation is sufficiently small or the mean value saturates and is unaffected by the addition of further circuits; the ideal value in the limit of a large number of random circuits is $\sim 0.85$.  The quantum volume is then defined as $2^{N_{max}}$ where $N_{max}$ is the largest number of qubits used in a circuit that passes the above test.

We perform a QV analysis on a 16 qubit IBM device with a reported QV of 32 (QV32) using both the expert settings and the Q-CTRL pipeline (as QV is a single-shot measure, we omit the post-processing measurement-error-mitigation component of the pipeline). We generate 300 random instances of QV circuits (using the QV method in Qiskit) and execute each with 1000 shots.  The execution of QV circuits is computationally intense and due to access constraints we elect to restrict our analysis to a moderate circuit count using a bootstrapping analysis of the mean cumulative heavy output probability for each value of circuit width.

\begin{figure}[t] 
\centering
\includegraphics[width=\columnwidth]{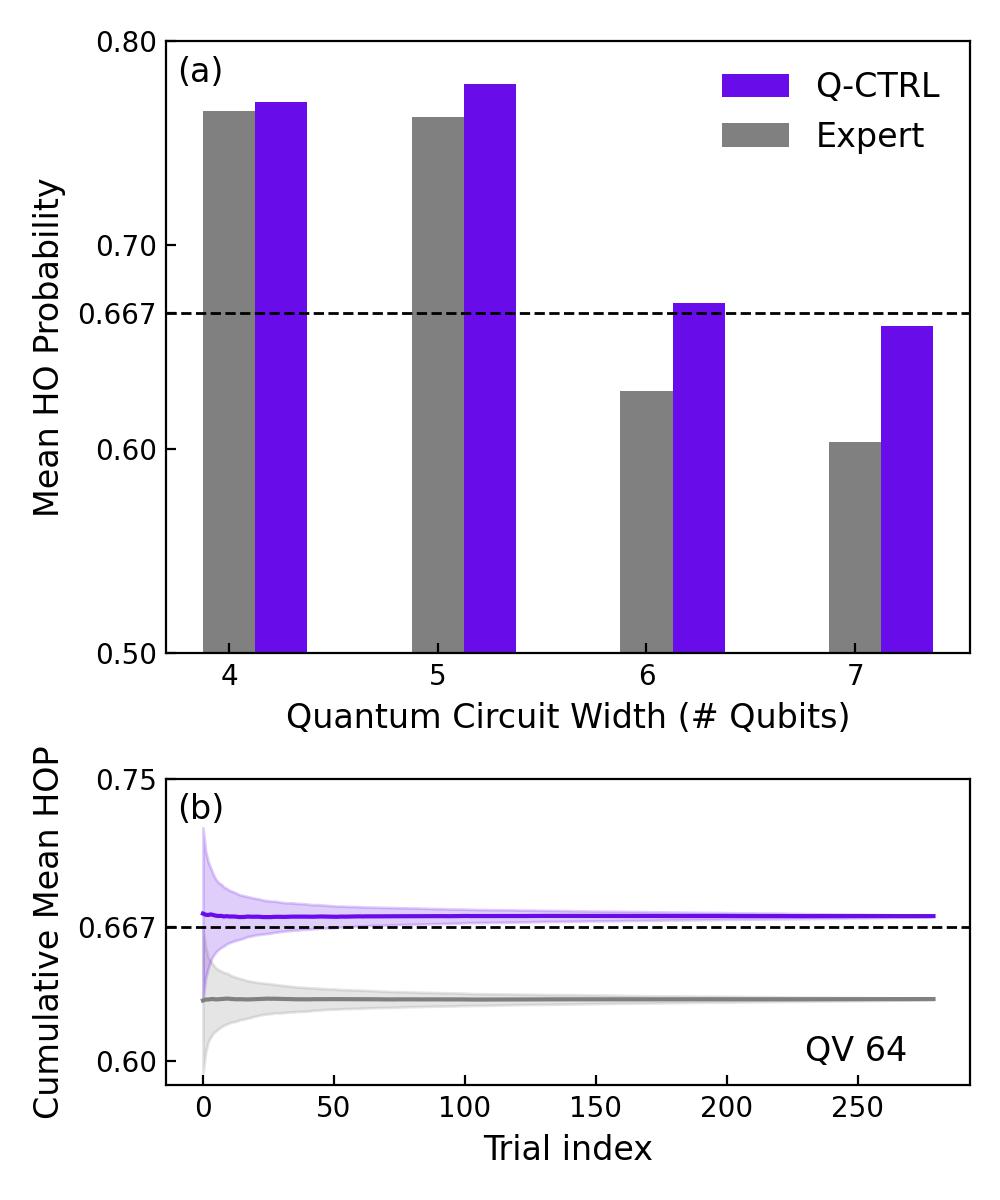}
\caption{A quantum volume experiment on a 16 qubit IBM device with reported QV of 32. In the experiment, we generate 300 random circuit instances and sample each 1000 times. We then calculate the mean heavy output probability over the 300 random circuits. (a) shows the mean HO probability vs. the number of qubits as obtained using both expert and Q-CTRL settings. (b) shows the cumulative sum and standard error as extracted from bootstrapping the 300 data points obtained in the 6Q experiment.}
\label{fig:QV}
\end{figure}

As seen in Fig.~\ref{fig:QV}a, the heavy output probability obtained using the Q-CTRL pipeline is consistently closer to the ideal 0.85 value.   The  Q-CTRL data is consistent with QV64 and is within $\sim1\%$ of achieving QV128, while the expert settings yield results consistent with QV32 as reported for this hardware system. 
Examining the cumulative mean HO for the 6 qubit case (Fig.~\ref{fig:QV}b), confirms that the value we measure is saturated to a value above the success threshold. Here shading indicates the range of mean values achieved for different random orderings of the trial averaging process in order to avoid systematic biases in the analysis.

\section{Discussion and outlook}
At present, the gap between anticipated performance extrapolated from proxy measures such as gate-level randomized benchmarking and actual algorithmic performance leaves substantial room for ongoing improvement. 
In our observations, this gap arises frequently from circuit-level error processes including classical and quantum crosstalk that require dedicated error reduction strategies. 
Our approach to this challenge has been to focus on deterministic techniques, leveraging physics-based knowledge of the dominant error processes gleaned by system identification experiments. 

The results presented above indicate that an autonomous pipeline leveraging deterministic error suppression strategies from the gate through to the compiler level can deliver large performance advantages across a range of algorithmic benchmarks and system-level proxy measures. 
Experiments on superconducting quantum computers with up to 16 qubits demonstrate up to $>1000\times$ enhancement of algorithmic success on shallow-depth algorithms relative to default configurations. 
As a corollary, we observed signatures of strong non-Markovian errors in these circuits which were efficiently suppressed near bounds set by $T_{1}$ processes via deterministic error suppression.  Benefits up to $>100\times$ enhancement persisted when compared against algorithmic execution including multiple expert-configured tools for error mitigation and suppression. 
In high-depth circuits we observe up to $7\times$ enhancement, again approaching limits imposed by incoherent errors. 
Our benchmarks also reveal the ability to improve the structural similarity of the experimental cost landscape in QAOA relative to an ideal-noise-free landscape by $163\times$. 
Combining this open-loop deterministic error-reducing pipeline with the execution of quantum error correction increases the average success of syndrome identification of an engineered error by $3.3\times$. 
Finally, we demonstrate enhancement of Quantum Volume on a 16 qubit device from $32\to 64$.

The approach we have demonstrated here, and the workflow encapsulated in the {\sl Fire Opal} software, possess multiple advantages relative to existing strategies for error reduction in quantum computers. 
First, the underlying error-suppression techniques used in algorithmic execution are \emph{deterministic}, requiring zero user overhead in terms of additional sampling or randomization;  this saves both wall-clock time and user cost when executing on hardware. 
Deterministic techniques for coherent averaging leveraging physics-based knowledge of the underlying hardware-noise sources are known to deliver superior performance relative to randomization strategies and require zero repetition, at the cost of including additional single-qubit gates (or more complex gate waveforms) in circuit execution. 
Nonetheless, for NISQ-era algorithms deterministic strategies are completely compatible with additional layers of sampling-based approaches such as zero-noise extrapolation or randomized compiling; deterministic error-suppression techniques must only constitute the final compiler pass in order to preserve the relevant physical frames for coherent averaging. 

Next, the {\sl Fire Opal} toolchain requires zero user configuration and is executed with a single command, obviating the need to manually include additional code blocks in circuit execution.
By contrast, existing tools typically require substantial configuration, making them inappropriate for users without detailed knowledge of the underlying physical mechanisms for error suppression. 
Finally, as our demonstrations validate, this strategy is applicable to both NISQ-era and fault-tolerant algorithmic execution, as the open-loop error-suppression strategies in use are fully compatible with the execution of quantum error correction protocols.  Deterministic error suppression helps reduce average hardware error rates below the fault-tolerant threshold, ultimately improving the encoding efficiency of QEC. In practice, we have seen that by virtue of having no requirement of additional sampling overhead, it is possible to suppress errors occurring during the QEC encoding and syndrome measurement processes in order to increase the likelihood of correctly identifying errors. 

Achieving sufficient \emph{absolute} algorithmic performance for practical applications will necessitate ongoing advances in hardware capability.  This primarily includes enhanced incoherent lifetimes ($T_{1}$) which currently pose restrictive bounds on both individual gate error rates and overall algorithmic performance for deep circuits.  An empirical observation in our demonstrations is that the magnitude of performance improvement due to the application of the error-suppression pipeline grows exponentially with qubit count.  We are therefore confident that as base hardware systems grow in scale and improve in performance that these approaches to deterministic error suppression in algorithmic execution will continue to yield fruit. 

\begin{acknowledgments}
The authors are grateful to all other colleagues at Q-CTRL whose technical, product engineering, and design work has supported the results presented in this paper. The authors acknowledge with gratitude IBM research for providing access to the hardware used in these experiments.
\end{acknowledgments}

\appendix

\section{Summary of approaches to error management in NISQ era devices}\label{Sec:AlternateStrategies}
In advance of the achievement of full-scale quantum error correction beyond the performance breakeven point, a range of techniques has been developed for the management of errors in NISQ-era hardware.  These approaches utilize different physical mechanisms but share the commonality of being \emph{open-loop} in that they are designed to reduce errors without using measurement feedback.

\begin{figure}[b!] 
\centering
\includegraphics[width=\linewidth]{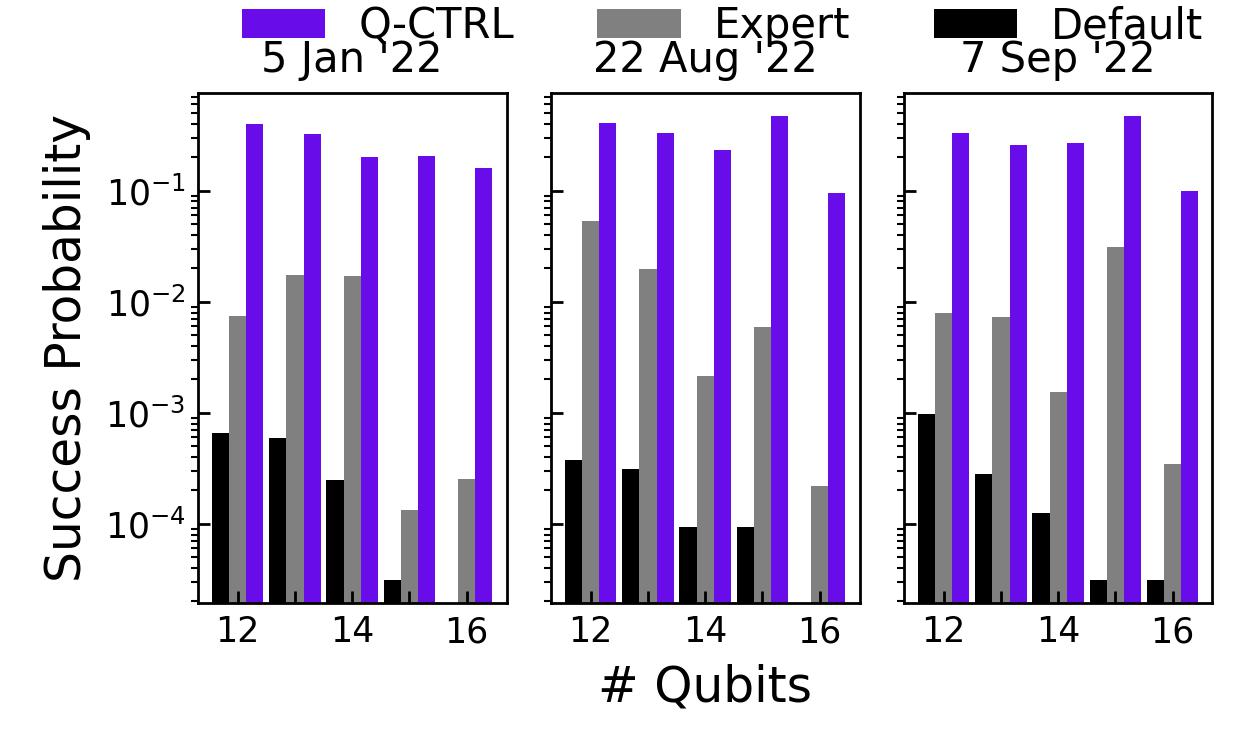}
\caption{Success probability (on a logarithmic scale) of the BV algorithm on three different dates (all different than the plot in the main text).  Data presented on a log scale.}
\label{fig:BV_stability}
\end{figure}

\begin{itemize}
    \item Circuit depth reduction (compilation) \cite{Sivarajah20, LiGushu18, Jones22, Maronese21}: The dominant approach to algorithmic performance enhancement is to employ mathematical identities to simplify quantum circuits, reducing gate counts and the duration of an executed circuit.  Challenges: typical compiler outputs are stochastic, requiring multiple passes to deliver a high-performance circuit;  run-to-run fluctuations in circuit characteristics and performance.
    \item Circuit layout optimization (compilation) \cite{Pedram16, Paler20, Murali19, Nation2023}: Due to the inhomogeneity of errors in typical devices, a compilation strategy may be ``noise aware'' in selecting a layout or qubit assignment which avoids faulty devices or minimizes use of known error-prone gate operations.  
    \item Dynamic decoupling for dephasing suppression \cite{tripathi2022, Santos05, Sekatski16, Pokharel18}: The physics of coherent averaging can be used to reduce dephasing on idling qubits in quantum algorithms via the addition of refocusing single-qubit pulses to the circuit.  Challenges: DD sequences are generally designed for single-qubit coherence, without consideration of circuit-level contextuality.
    \item Error-robust gate design:  Quantum logic gates may be designed with echo-like physics exploiting coherent averaging to cancel Hamiltonian noise terms during gate execution \cite{soare2014, carvalho2020errorrobust, Sheldon16, Sundaresan20, Patterson19}.  Challenges: individual gate-level optimization alone does not address circuit-level and contextual errors.
    \item Randomized compiling \cite{Wallman16, Hashim21, Cai19, Li17, Temme17, Endo18}: A circuit can be ``decorated'' with a randomly selected transformation applied by injected single qubit gates throughout the circuit.  This transformation may be tracked such that it requires no additional operations, but requires averaging over many different randomizations of the circuit.  A related approach called Probabilistic error cancellation~\cite{Minev_PEC} involves using random sampling to create a noise map which can then be deployed to invert the process in algorithmic execution.  Challenges: already for small numbers of qubits, requires user overhead of 10-20$\times$ circuit repetition to attain noise suppression, e.g., in Ref \cite{Hashim21} 50-100 randomizations per circuit were used in case of 4Q QFT. 
    \item Zero-noise extrapolation \cite{Temme17, Tiron20, Li17, Kandala19}: In attempting to improve the estimation of a quantum computer's output under noisy evolution it is possible to deliberately add noise systematically and subsequently extrapolate to the expected performance in the absence of noise. Challenges: This approach requires repeated execution of the same circuit with additional noise channels (\emph{e.g.} extended-duration gates).
\end{itemize}

\section{Stability of performance enhancement over time}\label{App:Stability}
Device performance varies both within a day and over longer periods~\cite{Carvalho2021}. In this section, we investigate the stability of the Q-CTRL performance. In Fig.~\ref{fig:BV_stability} we show the measured success probability of the Bernstein-Vazirani algorithm on three different dates (all different than for the data presented in the main text (Fig~\ref{fig:BV}). The Q-CTRL performance is consistently better than the expert and default comparisons on each day. There is also significantly less variation in the Q-CTRL performance across the three days. I.e, the expert configuration can vary up to an order of magnitude over different days. 

\section{Impact of different error suppression modules}\label{App:C}
In this section, we describe the different parts of the workflow and demonstrate their performance on real quantum devices.

\subsection{Front-end and error aware compilation}\label{Sec: compilation}
Quantum compilers are essential for finding more efficient ways to execute programs on quantum devices. A quantum compiler uses a ``compiler pass", consisting of a set of ``transformation passes", to handle circuit synthesis, optimization, and mapping to the device gateset and topology. Examples of transformation passes for optimization include: removing unnecessary gates, reducing circuit depth, and moving gates onto qubits with lower error rates. This simplification of the logic of the circuit is an integral first step in mitigating the impact of noise during program execution.

To compile the circuit optimally, however, a delicate balance between these circuit transformations must be found. For example, moving single-qubit gates onto qubits with low error rates requires the insertion of SWAP operations into the circuit. These SWAP gates increase circuit depth due to being made up of two-qubit gates (e.g., CNOT, CZ), which are also typically more prone to crosstalk such as $ZZ$ errors. An additional transformation pass could be used to minimize the number of SWAP operations, but this may effectively negate the benefit of the initial error-based mapping. It's these diminishing returns between each transformation that must be weighed for any given algorithm and device. 

As highlighted above, finding the optimal set of transformation passes is non-trivial and depends on both the device and the algorithm. There exist several open-source quantum compilers that can be used as a foundation to automate this procedure. 

\begin{figure}[t] 
\centering
\includegraphics[width=\linewidth]{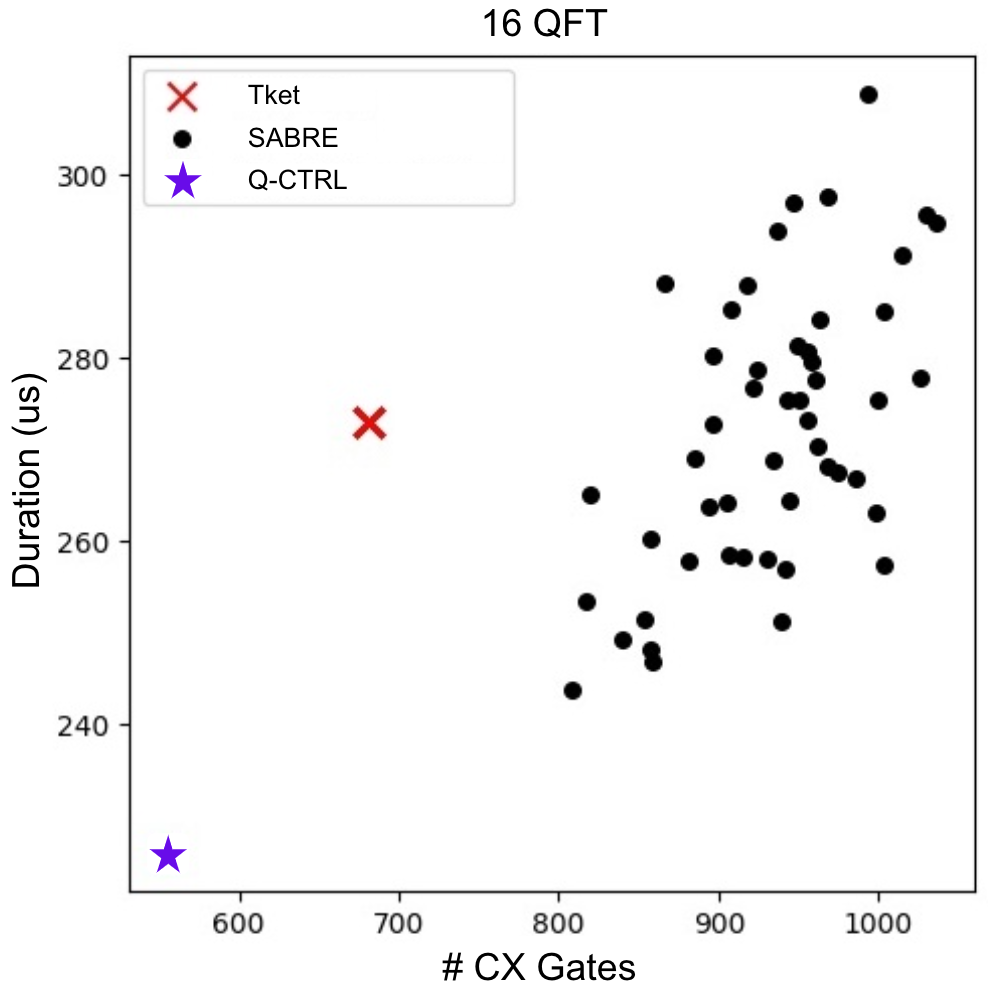}
\caption{We compare the different compilation methods using the 16-qubit QFT circuit. SABRE is a stochastic method, therefore, we present $50$ instances of the method. We use here the total number of CX gates and the total circuit duration as comparison features. As seen, the Q-CTRL method typically finds shorter circuits with a lower number of entangling gates. }
\label{fig:compilers}
\end{figure}

Our approach splits the effort into two parts. The first step is pure depth reduction and mapping to basis gates. As two-qubit gates (and SWAP gates in particular) are the main bottleneck, finding a minimal circuit depth for a given algorithm and device connectivity is crucial.
Once a minimal circuit is found (in terms of basis gates), there are multiple qubits layouts that may host such a circuit. Layout optimization considers information about various aspects that affect the circuit performance, e.g., gate and measurement errors, classical and quantum crosstalk, and decoherence times. In Fig.~\ref{fig:compilers}, we compare two open-source compilation methods, SABRE\cite{Li_Sabre2018} and Tket\cite{Sivarajah_tket_2020}, with our hybrid approach for 16-qubit QFT algorithm. To elucidate the variation of the SABRE method, we present the results of 50 independent instances of the method. Our method consistently finds circuits with shorter durations and a lower number of entangling gates. Subsequently, our AI-powered layout selection protocol addresses the inhomogeneity of the device by avoiding low-quality qubits and gates.

\subsection{Circuit level error suppression via dynamic decoupling}\label{Section:DD}
Dynamical decoupling (DD) is an open-loop quantum control technique aimed at suppressing idling errors. In its simplest form, DD is implemented by periodic sequences of quantum bit flips, whose net effect is to approximately cancel unwanted couplings. Most traditional DD protocols are designed to reduce global dephasing ($T_2$ processes), but different schemes exist for designing DD protocols for specific error-reducing tasks \cite{Pokharel18, tripathi2022}. For instance, crosstalk errors due to spurious $ZZ$ coupling can be mitigated by a DD protocol. Yet, protocols that mitigate $ZZ$ errors can change dramatically from one algorithm to another and from one device to another, depending on the device topology, connectivity, and ability to parallelize the different layers of the algorithm.

Our automated scheme takes as input: a quantum circuit, backend characterization, and device topology. It returns a new circuit, in which an optimal DD protocol is embedded. We optimize the trade-off between the number of $X$ gates (providing robustness to higher frequency noise processes) with the inherent gate error of the $X$ gates. 

\begin{figure*}[t] 
\centering
\includegraphics[width=\linewidth]{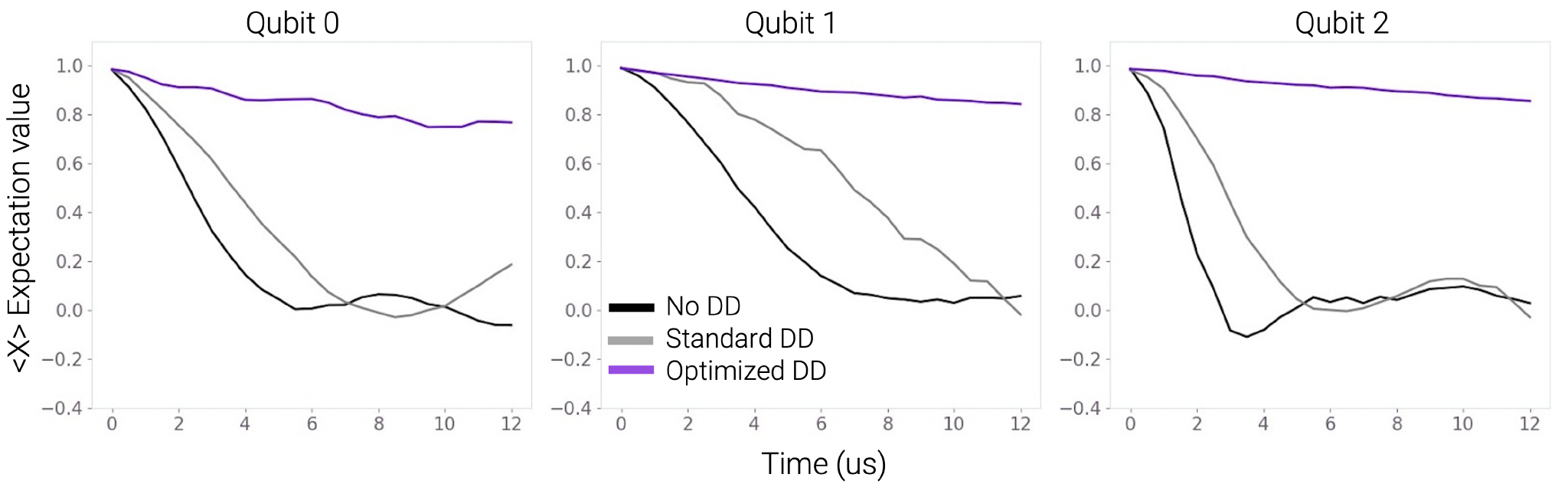}
\caption{A Ramsey-like experiment where three coupled qubits are prepared at the $|+\rangle$ state (i.e., $\langle X\rangle = 1$). The state should remain unchanged under zero error conditions. Optimized DD significantly reduces the idling error when compared to the standard DD protocol.}
\label{fig:free_evol}
\end{figure*}

For a pair of coupled qubits, a $ZZ$ idling error exists on top of the $T_2$ error. A bit flip on a single qubit inverts the Hamiltonian evolution due to the $ZZ$ coupling, while a simultaneous application of bit flips on each qubit leaves the $ZZ$ coupling unchanged. To mitigate $ZZ$-error, the timings of bit flip on the two qubits need to be staggered and correlated. 
In Fig.~\ref{fig:free_evol} we show a Ramsey experiment where three coupled qubits are prepared in the $|+\rangle$ state (i.e., $\langle X\rangle = 1$). The qubit states were measured after different idling times in the absence of DD, in the presence of an uncorrelated (or standard) DD sequence, or in the presence of a correlated (or optimized) DD sequence. The non-correlated standard DD reduces the idling errors due to $T_2$ processes but it cannot account for errors due to unwanted $ZZ$ coupling.

Finding a multi-qubit correlated DD sequence for fully idle qubits is a relatively easy task that only requires careful timing of the different bit flips.
However, finding the optimal DD embedding in an arbitrary algorithm has no clear solution. The presence of gates introduces a different context at different layers of the algorithm. This renders the embedding task into a contextual optimization problem where the correlated DD sequences are locally adjusted to the context at any point in time.

\begin{figure*}[t!] 
\centering
\includegraphics[width=\linewidth]{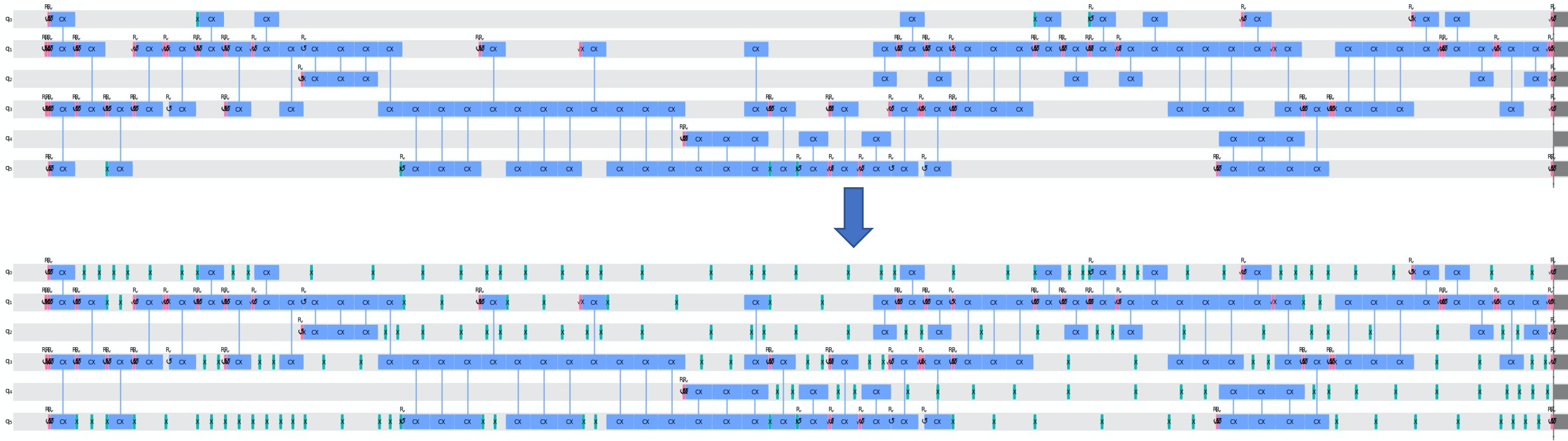}
\caption{An example of the DD embedding. The input circuit, on the top, is a compiled circuit and the output circuit of our automated scheme, on the bottom, is the original circuit with the optimal DD structure embedded in it.}
\label{fig:input_output_example}
\end{figure*}

Our method is designed to cancel as much as possible the $T_2$ and $ZZ$ errors (typically, full cancellation is not possible) while keeping the number of $X$ gates as low as possible. Extremely dense DD schemes may be more harmful than useful, due to the individual errors of each $X$ gate and heating effects. 
See Fig.~\ref{fig:input_output_example} for an example of input and output circuits.
The embedding procedure is hardware agnostic and completely automated. Given a compiled circuit, device topology, and backend data, it returns the circuit with the optimal DD embedded in it.

\subsection{System wide gate optimization}\label{Sec: calib}

We employ an AI-driven autonomous gate-optimization protocol that designs low-level analog definitions of gate waveforms through closed-loop optimization. This strategy requires no prior knowledge of the device model or its error processes.

We use multiple simulated annealing agents in parallel to optimize the estimated gate infidelity for each qubit pair. In each step of the optimization, each agent uses measurements obtained from multiple test circuits in order to estimate the operation's infidelity with respect to the ideal outcome (given a perfect target gate). The test circuits are chosen in a way that sequentially amplifies the gate errors. An effective measure of error-per-gate is extracted by applying a linear fit to the infidelity of the different circuits as a function of the noise level, which provides a measure of the gate error in the low error limit (as cross-validated using iRB). 

\begin{figure}[h] 
\centering
\includegraphics[width=\linewidth]{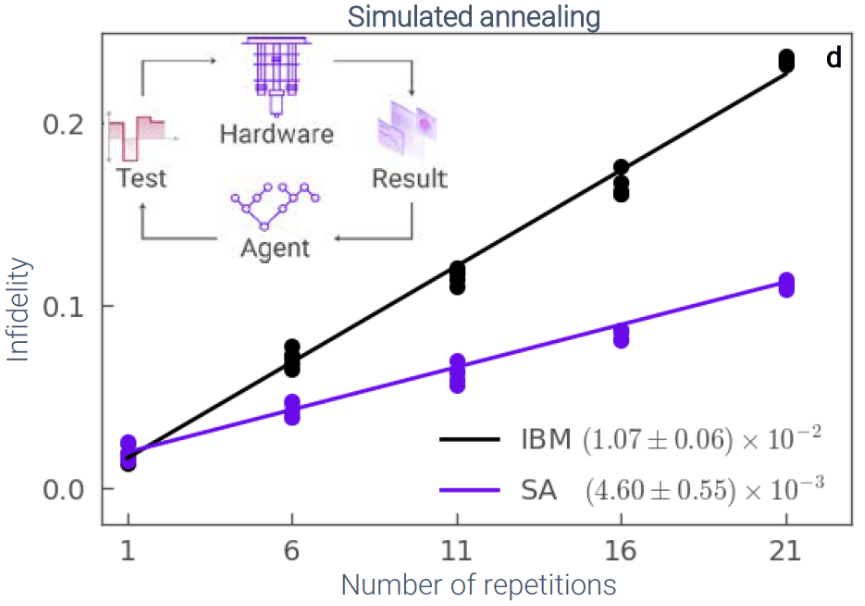}
\caption{(Figure adapted from Ref.~\cite{Baum2021}) A comparison between a default CNOT gate and an optimized gate found by a simulated annealing optimizer. Each gate is applied $N$ times on two different initial states and the infidelity with respect to the ideal state was calculated by performing a full-state tomography. The gate errors are extracted from the slope of the infidelity, showing a $\sim 2.3\cross$ improvement compared to the default.}
\label{fig:gate_opt}
\end{figure}

The above procedure constitutes a cost function that assigns a quality measure to specific gate implementations. Our automated optimization procedures are designed to interact with the system in order to produce better and better gate implementations. More details can be found in Ref.~\cite{Baum2021}. In Fig.~\ref{fig:gate_opt} we show a comparison between the default CNOT gate and an optimized gate found by a simulated annealing optimizer. In this comparison, each gate is applied $N$ times on two different initial states and the infidelity with respect to the ideal state was calculated by performing a full-state tomography.

\begin{figure}[h] 
\centering
\includegraphics[width=\linewidth]{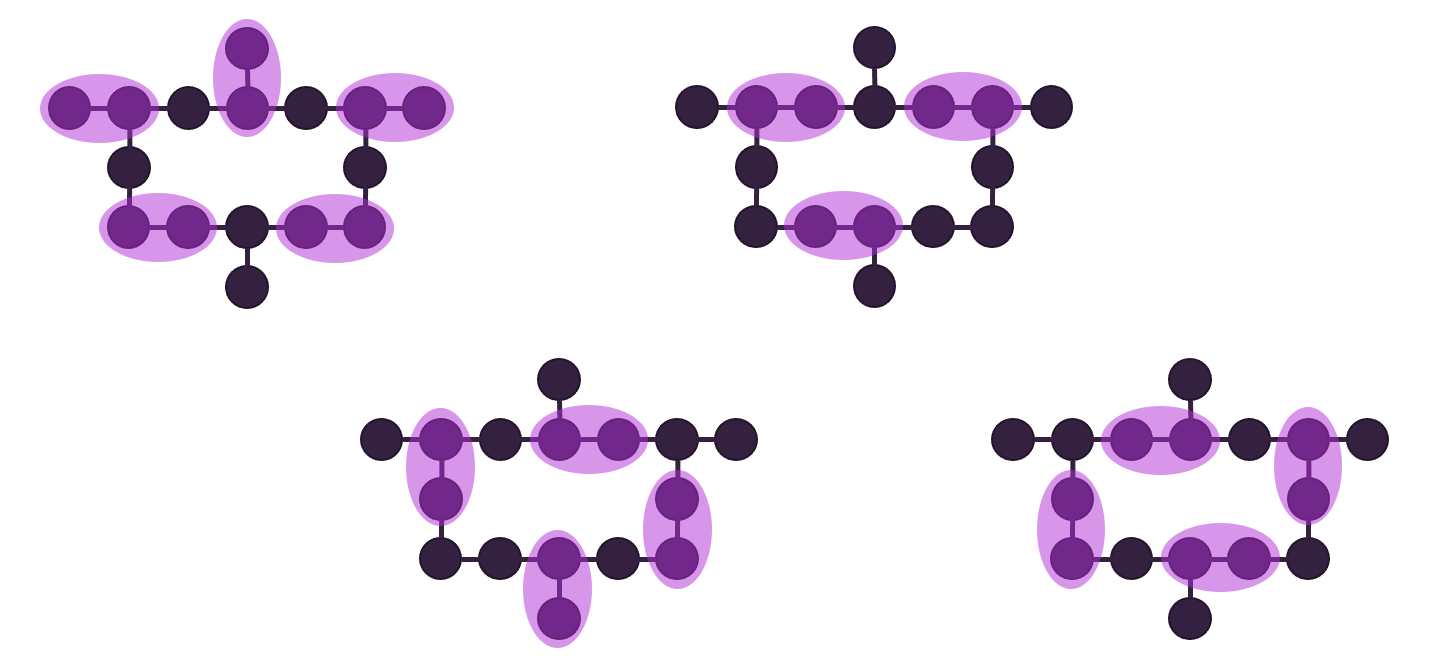}
\caption{Full device calibration can be done in parallel requiring a fixed number of steps depending on the device connectivity. Heavy-hex topology can be tuned up in only four steps irrespective of the device size.}
\label{fig:calibration_grp}
\end{figure}

We parallelize multiple agents across the device in order to simultaneously optimize all two-qubit gates. The process structure is dictated by the connectivity of the device; a qubit cannot be shared between pairs being optimized simultaneously. Furthermore, to avoid cross-talk effects, we avoid a parallel optimization of neighboring pairs. Given the IBM device connectivity, we parse the device into four disconnected parallel blocks as depicted in Fig.~\ref{fig:calibration_grp} and execute parallel optimizations over all gates within a block.  The process is repeated over each block until all two-qubit gates are optimized. 

\subsection{Measurement error mitigation}\label{Sec: mit}
\begin{figure}[t!] 
\centering
\includegraphics[width=\linewidth]{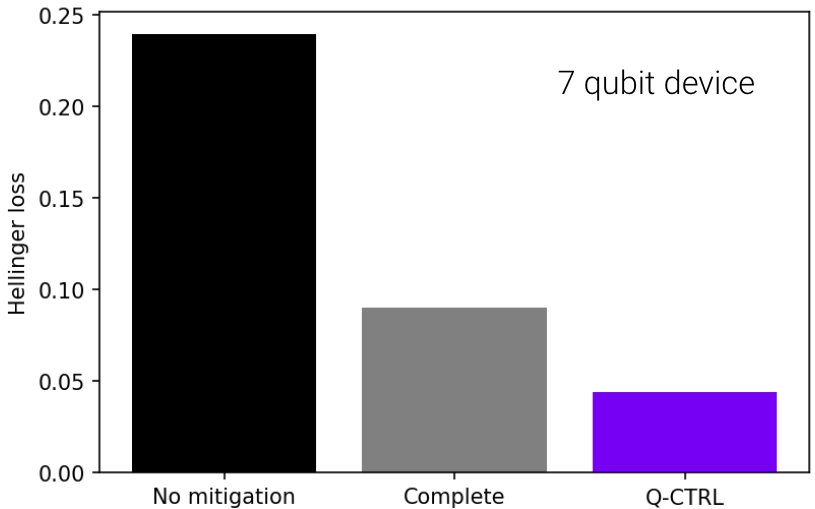}
\caption{Comparison of readout error, denoted by the Hellinger loss, averaged over 2000 random probability distributions on a 7-qubit quantum computer. The complete method includes a full calculation of the confusion matrix which requires $2^7$ experiments to construct the filter.}
\label{fig:NN_comp}
\end{figure}

We implement a custom measurement error mitigation protocol designed to suppress readout errors in the hardware. The number of calibrations required in the protocol scales sub-linearly with the number of qubits and as shown in Figure~\ref{fig:NN_comp}, gives better mitigation than complete measurement mitigation protocol. We discuss the details of the two protocols below.

We represent the error in readout assignment for a single qubit with a confusion matrix $C_{ij} = P[i|j]$, i.e., the conditional probability of assigning $\left|i\right>$ to a measured state $\left|j\right>$. For a single qubit, given a measured vector of probabilities $\vec{P}_{meas}$, one can obtain the ideal probability distribution by minimizing the quantity $|C\vec{P}_{ideal} - \vec{P}_{meas}|^2$, under the constraint, that $\vec{P}_{ideal}$ is a valid probability vector (non-negative elements that sum to one). However, this method is prone to several scalability issues- 

\begin{itemize}
    \item The size of the confusion matrix for a multi-qubit system ($2^{N_q}\cross2^{N_q}$) scales exponentially in the number of qubits ($N_q$). For $N_q=42$, storing $C$ \cite{Nation2022} in a sparse format requires 580PiB which is 120-times more than that available on Fugaku \cite{Fugaku} the fastest supercomputer.
    \item Estimating $C$ requires $2^{N_q}$ calibration experiments.
    \item Since the measured distribution itself has $2^{N_q}$ different possible outcomes, sampling only $M$ times with $M\ll2^{N_q}$ produces a poor estimation of the confusion matrix.
    \item Solving the constrained minimization problem for more than a handful of qubits is often ill-defined and is a time-consuming process.
\end{itemize}
We overcome these challenges by dividing a quantum device into smaller groups of qubits which can then be efficiently corrected via tensor algebra. The groups are automatically chosen in a way that minimizes inter-group correlations among the measurements based on the topology of the given quantum computer.
It is essential to ensure that all the groups combined span the whole quantum computer. 

Let $N_g$ be the number of groups and $\{G_i\}$ for $i \in \{k \in \mathbb{N}:k \leq N_g\}$ denote the set of groups such that $\bigcup\limits_{i=1}^{N_g} G_{i} = [N_q]$. The confusion matrices corresponding to each group in $\{G_i\}$ are $\{C_i\}$ and the size of each $C_i$ matrix is given by $2^{|G_i|}\cross2^{|G_i|}$.

The number of calibration experiments required for our method is given by $2^{\max\limits_{i}(|G_i|)}$. As we limit the maximal number of qubits in each group, the number of calibration experiments does not scale with the system size.

The output from tensored mitigation above serves as an input to a neural network. The goal of the neural network is to restore correlations neglected (at the boundaries of the groups) and account for nonlinear effects and contextual errors that are not captured in the confusion matrix formalism. Since we are looking for small modifications to the identity transformation, we use a ResNet-like architecture for our neural network. The number of tunable parameters in our neural network scales sub-linearly with the number of qubits ($N_q$). This approach minimizes the amount of training data required.

%
\end{document}